\documentclass[12pt,epsfig,color]{article}
\usepackage{amsmath}
\usepackage{amsfonts}
\usepackage{amsmath}
\usepackage{slashed}
\usepackage{amsxtra}
\usepackage{amstext}
\usepackage{amssymb}
\usepackage{color}
\usepackage{float}
\usepackage{graphicx}
\usepackage{subcaption}
\numberwithin{equation}{section}
\usepackage[dvips]{epsfig}
\usepackage{appendix}
\usepackage{youngtab}
\newcommand{\be}{\begin{equation}}
\newcommand{\ee}{\end{equation}}
\newcommand{\beq}{\begin{equation}}
\newcommand{\eeq}{\end{equation}}
\newcommand{\ba}{\begin{eqnarray}}
\newcommand{\ea}{\end{eqnarray}}
\newcommand{\bea}{\begin{eqnarray}}
\newcommand{\eea}{\end{eqnarray}}

\begin{document}
\baselineskip=15.5pt \pagestyle{plain} \setcounter{page}{1}
%
\begin{titlepage}

\vskip 0.8cm

\begin{center}
%
%
%
%
%

{\Large \bf  Higher-twist fermionic operators and DIS structure functions from the AdS/CFT duality}

\vskip 1.cm

{\large {{\bf David Jorrin}{\footnote{\tt jorrin@fisica.unlp.edu.ar}} {\bf and  Martin
Schvellinger}{\footnote{\tt martin@fisica.unlp.edu.ar}}}}

\vskip 1.cm

{\it Instituto de F\'{\i}sica La Plata-UNLP-CONICET. \\
Boulevard 113 e 63 y 64, (1900) La Plata, Buenos Aires, Argentina \\
and \\
Departamento de F\'{\i}sica, Facultad de Ciencias Exactas,
Universidad Nacional de La Plata. \\
Calle 49 y 115, C.C. 67, (1900) La Plata, Buenos Aires, Argentina.}

\vspace{1.cm}

{\bf Abstract}

\end{center}

\vspace{1.cm}

The role of local higher-twist ($\tau > 3$) spin-1/2 fermionic operators of the strongly coupled ${\cal {N}}=4$ supersymmetric Yang-Mills theory on the symmetric and antisymmetric deep inelastic scattering (DIS) structure functions is investigated. The calculations are carried out in terms of the duality between ${\cal {N}}=4$ SYM theory and type IIB supergravity on AdS$_5 \times S^5$. Particularly, we explicitly obtain the structure functions for single-trace spin-1/2 fermionic operators in the {\bf 20$^*$} and {\bf 60$^*$} irreducible representations of $SU(4)_R$, corresponding to twists 4 and 5, respectively. We also calculate the contributions of other single-trace spin-1/2 fermionic operators in the {\bf 4}, {\bf 20} and {\bf 60} irreducible representations of $SU(4)_R$. New important effects are found  in comparison with the minimal twist ($\tau = 3$) case, and they are studied thoroughly.

\noindent

\end{titlepage}

\newpage

{\small \tableofcontents}

\newpage


%
\section{Introduction}
%

Deep inelastic scattering (DIS) cross sections of charged leptons by hadrons are expressed as the contraction of a leptonic tensor with a hadronic one. The leptonic tensor is easily obtained from QED. The problem lies within the calculation of the hadronic tensor, which is given in terms of the two-point function of electromagnetic currents within the hadron, where strong coupling effects become important. In the operator product expansion (OPE) of two electromagnetic currents inside a hadron there are several kinds of contributions from different SYM theory operators, which in certain parametric domains can be relevant depending on the virtual-photon momentum $q$, the coupling $g_{SYM}$, the Bjorken parameter $x$, and the number of color degrees of freedom $N$. Certain properties of the hadronic tensor as well as relations among different structure functions, such as the Callan-Gross relation and generalizations of it, are valid for different gauge field theories, at least within the same parametric regimes of $q$, $g_{SYM}$, $x$, and $N$. In particular, this behavior has been found in the framework of the gauge/string theory duality \cite{Maldacena:1997re,Gubser:1998bc,Witten:1998qj} in diverse situations for the strongly coupled regime of different gauge theories, starting from the pioneering work by Polchinski and Strassler \cite{Polchinski:2002jw}. Structure functions of spin-1/2 hadrons have been investigated in this context in \cite{Polchinski:2002jw,Gao:2009ze,Gao:2010qk,BallonBayona:2007qr,BallonBayona:2009uy,Kovensky:2018xxa,Jorrin:2020cil}. These techniques have been also applied to the study of the structure functions of scalar and vector mesons from Dp-brane systems with flavor branes preserving some supersymmetries as in the D3D7-brane model \cite{Kruczenski:2003be}, or breaking supersymmetry completely as in the Sakai-Sugimoto model \cite{Sakai:2004cn} and in the D4D6 anti-D6-brane model \cite{Kruczenski:2003uq}, which have been considered in \cite{BallonBayona:2008zi,BallonBayona:2010ae,Bayona:2011xj,Koile:2011aa,Koile:2013hba,Koile:2014vca,Koile:2015qsa}. Also, $1/N$ corrections have been investigated in this context \cite{Jorrin:2016rbx,Kovensky:2016ryy,Kovensky:2018gif}. In addition, very important holographic Pomeron techniques have been developed and applied to different models derived from both type IIA and type IIB superstring theories \cite{Polchinski:2002jw,Brower:2007xg,Brower:2007qh,Cornalba:2006xm,Cornalba:2007zb,Costa:2012fw,Watanabe:2012uc,Costa:2013uia,Koile:2014vca,Koile:2015qsa,Kovensky:2018gif,Kovensky:2017oqs}. Another interesting related aspect is the DIS off a strongly coupled ${\cal {N}}=4$ SYM plasma  \cite{Hatta:2007cs}, as well as its corrections within the strong coupling expansion which have been obtained in \cite{Hassanain:2009xw} from $\alpha'^3$ string theory corrections to the type IIB supergravity action \cite{Paulos:2008tn}.

For the electromagnetic DIS let us consider an incident polarised spin-1/2 hadron, with four-momentum $P^\mu$, mass $M$, and a spin vector $S^\mu$. The corresponding hadronic tensor can be written as
\begin{equation}
    W_{\mu \nu}= W^{(S)}_{\mu \nu}(q,P)+i \ W^{(A)}_{\mu \nu}(q,P,s) \, ,
\end{equation}
which is expressed in terms of the Bjorken variable defined as $x=-q^2/(2 P \cdot q)$. The DIS limit corresponds to $q\to \infty$, while $x$ is kept fixed.  The hadronic tensor can writen in terms of the structure functions as follows   \cite{Anselmino:1994gn,Lampe:1998eu}
\begin{eqnarray}
 W^{(S)}_{\mu \nu}&=&  \left(\eta_{\mu\nu} - \frac{q_\mu q_\nu}{q^2}\right)\left[F_1(x,q^2)+\frac{M \ S \cdot q}{2 P \cdot q} g_5(x,q^2) \right] \nonumber\\
&& - \frac{1}{P\cdot q} \left(P_\mu - \frac{P \cdot q}{q^2} q_\mu \right)\left(P_\nu - \frac{P \cdot q}{q^2} q_\nu \right) \left[F_2(x,q^2) +\frac{M \ S \cdot q}{P\cdot q} g_4(x,q^2) \right] \nonumber\\
&&-\frac{M}{2 P\cdot q} \left[ \left(P_\mu - \frac{P \cdot q}{q^2} q_\mu \right)\left(S_{\nu}-\frac{S \cdot q}{P\cdot q} P_{\nu} \right)+\left( P_\nu - \frac{P \cdot q}{q^2} q_\nu \right)\left(S_{\mu}-\frac{S \cdot q}{P\cdot q} P_{\mu} \right)\right] \nonumber\\
&& g_3(x,q^2) \, ,
\end{eqnarray}
\begin{eqnarray}
 W^{(A)}_{\mu \nu}&=&- \frac{M \ \epsilon_{\mu \nu \rho \sigma } \ q^{\rho}}{P\cdot q} \left(S^{\sigma}g_1(x,q^2) +\left[ S^{\sigma}- \frac{S\cdot q}{P \cdot q} P^{\sigma} \right]g_2(x,q^2)  \right)- \frac{\epsilon_{\mu \nu \rho \sigma} q^{\rho} P^{\sigma}}{2 P \cdot q} F_3(x,q^2) \, . \nonumber \\
\end{eqnarray}
where we have separated the hadronic tensor into its symmetric and antisymmetric parts, and also we have used the metric defined as $\eta_{\mu\nu}=\text{diag}(-1,1,1,1)$. In addition, one can define another tensor $T^{\mu\nu}$, related to the forward Compton scattering, as the expectation value of the time-ordered product of two electromagnetic currents inside the hadron, 
\begin{equation}
T^{\mu \nu} \equiv  i \int d^4 \xi \ e^{i q \cdot \xi} \  \langle P ,{\cal{Q}}, S|{\hat{T}} \{J^{\mu}(\xi) J^{\nu}(0)\}|P, {\cal{Q}}, S\rangle \, . \label{Ttensor}
\end{equation}
Its imaginary part can be expressed as a sum over intermediate states that we call ${\cal {X}}$,
\begin{eqnarray}
\text{Im}(T^{\mu\nu})&=&2 \pi^2 \sum_{{\cal {X}}} \delta(M_{\cal {X}}^2+(P+q)^2) \langle P ,{\cal{Q}},S| J^{\nu}(0)  | P_{\cal {X}}, {\cal{Q}},S\rangle\langle P_{\cal {X}} ,{\cal{Q}},S| J^{\mu}(0)  | P, {\cal{Q}},S\rangle \label{optical-teo} \, . \nonumber \\
\end{eqnarray}
In terms of the optical theorem we have: 
\begin{eqnarray}
 W^{(S)}_{\mu \nu}=2 \pi \  \text{Im}(T^{(S)}_{\mu\nu})\, ,\ \ \  \ \ \ \ \ \ \ \  W^{(A)}_{\mu \nu}=2 \pi \ \text{Im}(T^{(A)}_{\mu\nu})\, .  \label{TW}
\end{eqnarray}

In the case of the planar limit of the strongly coupled ${\cal {N}}=4$ SYM theory with gauge group $SU(N)$, when $1 \ll \lambda_{SYM} \ll N$ one can explicitly calculate the hadronic tensor from its string theory dual description, given in terms of type IIB superstring theory on AdS$_5 \times S^5$ in the $\alpha' \rightarrow 0$ limit, {\it i.e.} type IIB supergravity, including an IR cut-off $\Lambda$ in order to account for color confinement \cite{Polchinski:2002jw}. In particular, when the Bjorken variable is in the $\lambda_{SYM}^{-1/2} \ll x < 1$ regime (where the 't Hooft coupling is $\lambda_{SYM} \equiv g^2_{SYM} N$), only type IIB supergravity fields are relevant for the holographic dual calculation of properties related to the DIS. In that parametric region the OPE of the two electromagnetic currents inside the hadron is dominated by double-trace operators obtained as the product of two protected single-trace operators. There is a factorization in terms of $(\Lambda^2/q^2)^{\tau-1}$. The twist is defined as $\tau = \Delta - s$, for an operator with scaling dimension $\Delta$ and spin $s$. In a previous paper \cite{Jorrin:2020cil} we have considered single-trace spin-1/2 fermionic operators with $\tau=3$ which belong to the {\bf 4$^*$} irreducible representation of $SU(4)_R$. For that purpose firstly we have derived the corresponding terms in the effective five-dimensional supergravity action containing the coupling of two dilatino modes with a massless vector field. We have done it from the dimensional reduction of type IIB supergravity on $S^5$. Those terms in the five-dimensional action, which we briefly discuss in Section 3 of the present work, are the minimal coupling and two Pauli terms, one of which connects the same incoming and intermediate states (in the forward Compton scattering related to the DIS process via the optical theorem) and a second one which allows for certain different intermediate states that we study in detail. In \cite{Jorrin:2020cil} we have shown that for $\tau=3$ spin-1/2 fermionic operators the effects due to Pauli terms account for about 90$\%$ of each structure function, thus they play a very important role in the DIS process of $SU(N)$ ${\cal {N}}=4$ SYM theory at strong coupling in the planar limit.

In the present work we investigate the contributions given by local single-trace higher-twist ($\tau > 3$) spin-1/2 fermionic operators of the strongly coupled $SU(N)$ ${\cal {N}}=4$ SYM theory on both the symmetric and the antisymmetric structure functions of a polarized spin-1/2 hadron. We consider the large $N$ limit. We work within the supergravity parametric domain, thus we consider the spontaneous compactification of type IIB supergravity on $S^5$. We focus on the structure functions related to twist $\tau=k+3$ spin-1/2 fermionic operators of the type ${\cal {O}}_k^{I_k, (6)}$ defined in Section 2. Our special interest is in the cases of twists 4 and 5, corresponding to $k=$1 and 2, respectively. In the calculation we also discuss the effect of the ${\cal {O}}_k^{I_k, (13)}$ single-trace spin-1/2 fermionic operators which, by virtue of the selection rules we found, also appear as possible final states in the DIS process we consider. It is interesting to emphasize that for single-trace higher-twist spin-1/2 operators there are important new effects that we investigate in this work. One of such effects comes from the fact that as $k$ increases the dimension of the irreducible representation of $SU(4)_R$ increases substantially leading to a large number of Kaluza-Klein dilatino modes contributing from the supergravity side. For instance, for $\tau=4$ there are 20 Kaluza-Klein modes related to the type IIB supergravity dilatino modes on AdS$_5 \times S^5$ which contribute to the calculation. Things become even much more complicated for $\tau=5$ where there are 60 spinors contributing to the calculation of the structure functions. In Section 2 we discuss the relation between SYM operators and Kaluza-Klein dilatino modes in each case. On the other hand, there are new additional terms providing relevant contributions coming from the fact that for $\tau>3$ the selection rule $\lambda_k \rightarrow \lambda_{k-1}$ with $k>0$ now plays a significant role. These contributions are not sub-leading in comparison with the contributions that appear for $\tau=3$ spin-1/2 fermionic operators.  Therefore, it is worth to investigate the effect of all these new contributions altogether on the hadronic tensor of spin-1/2 fermions. We will carry out a detailed calculation of the referred effects. This is very interesting because it allows us to understand better how the supergravity dual description accounts for the way the momentum fragmentation and evolution occur in the planar limit of the strongly coupled quantum field theory within the $\lambda_{SYM}^{-1/2} \ll x < 1$ range for spin-1/2 fermionic operators of ${\cal {N}}=4$ SYM theory.

The structure of this work is as follows. In Section 2 we describe the relation between single-trace spin-1/2 fermionic SYM theory operators and the Kaluza-Klein field modes obtained by considering the dimensional reduction of type IIB supergravity on $S^5$. In Section 3 we develop the dual type IIB supergravity calculation of the structure functions for the mentioned operators for twist 4 in Section 3.1 and for twist 5 in Section 3.2. In Section 4 we analyse our results and present the conclusions. There are in addition several appendices containing certain important details of the calculations.

%
\section{Spin-1/2 fermionic operators of ${\cal {N}}=4$ SYM and type IIB supergravity fields}
%

The ${\cal {N}}=4$ SYM gauge supermultiplet contains four left Weyl fermions which we label as $\lambda_{{\cal {N}}=4}$ (we use this notation to distinguish it from $\hat{\lambda}$ which represents the ten-dimensional dilatino field of type IIB supergravity). There are also $X_j$ real scalars with $j=1, \cdots , 6$, and $F_+$ labels the self-dual two-form field strength associated with the $SU(N)$ gauge field. All these fields transform in the adjoint representation of the gauge group $SU(N)$.

We focus on the structure functions corresponding to local twist $\tau=k+3$ spin-1/2 fermionic operators of the form ${\cal {O}}_k^{I_k, (6)}(x) = C^{I_k, (6)}_{i_1 \dots i_k} \, \text{Tr}(F_+ \lambda_{{\cal {N}}=4} X_{i_1} \dots X_{i_k})(x)$ where $i_j$ are indices corresponding to the 6 real scalars of the ${\cal {N}}=4$ SYM gauge supermultiplet. In addition, the integer $I_k$ runs from 1 to the dimension of the irreducible representation of $SU(4)_R$. These operators transform in the $[1, k, 0]$ irreducible representation of the $R$-symmetry group of the ${\cal {N}}=4$ SYM theory, being $k \ge 0$ (see for instance \cite{Aharony:1999ti,DHoker:2002nbb} and references therein). The case for $k=0$ has been investigated in detail in \cite{Jorrin:2020cil}. In that situation there are just 4 operators of the form ${\cal {O}}_{k=0}^{I_0, (6)}(x) = C^{I_0, (6)} \text{Tr}(F_+ \lambda_{{\cal {N}}=4})(x)$, corresponding to $\tau=3$, which are in the {\bf 4$^*$} irreducible representation of $SU(4)_R$. As the number of scalar fields becomes larger the complexity of the calculation increases dramatically since the dimension of the corresponding irreducible representation grows with $k=0, 1, 2, 3, 4, \cdots$ as 4, 20, 60, 140, 280, $\cdots$. This means that in terms of the AdS/CFT duality one has to deal with an increasing number of operators on the gauge theory side, and also with the same number of Kaluza-Klein dilatino modes on the type IIB supergravity side. For this reason, and in order to show explicitly the new effects we find for higher-twist operators we only carry out the explicit calculations of the hadronic tensor in the cases of twist-4 and twist-5 spin-1/2 fermionic operators. For higher-twist operators ${\cal {O}}_k^{I_k, (6)}(x)$ the same method can be applied.

In order to calculate the dimension of an irreducible representation of $SU(4)_R$ it is useful to consider in general the irreducible representations of the $su(n)$ algebra of the $SU(n)$ Lie group. Recall that a simple Lie algebra has a Cartan sub-algebra of rank $r$ and an associated root space spanned in a basis given by the corresponding simple roots, $\alpha_i$, with $i=1, \cdots, r$. There is also a reciprocal basis of vectors $\beta^j$, $j=1, \cdots, r$. An irreducible representation of the Lie algebra can be described in terms of its highest weight vector $V=\sum_{j=1}^{r} m_j \beta^j$, where $m_j$ are the Dynkin integers labelling the different irreducible representations $[m_1, m_2, \dots, m_r]$ of $su(n)$. The dimension of $[m_1, m_2, \dots, m_r]$ can be calculated very easily by associating a Young diagram with that representation as follows. One must construct a Young diagram with $m_j$ columns of length $j$ (the length is given by the number of single boxes in that column). The relation between $n$ and $r$ is $n=r+1$, thus for the Lie group $SU(4)_R$ the rank of its Cartan sub-algebra is $r=3$, {\it i.e.} there are only three simple roots, therefore the irreducible representations of  $SU(4)_R$ can be labelled by three Dynkin integers $[r_1, r_2, r_3]$\footnote{Notice that we have now switched to the standard notation by calling $r_j$ to the Dynkin labels of the irreducible representation of the $SU(4)_R$ Lie group, {\it i.e.} $[m_1, m_2, m_3] \equiv [r_1, r_2, r_3]$.}. In particular, for the single-trace spin-1/2 fermionic operators ${\cal {O}}_k^{I_k, (6)}(x)$ the irreducible representations of $SU(4)_R$ are $[1, k, 0]$ with $k \ge 0$. Also, these operators transform in the $(1/2,0)$ representation of the algebra of $SU(2) \times SU(2)$ which is isomorphic to the complexified algebra of the Lorentz group $SO(1,3)$, while their conformal dimensions are $\Delta=k+\frac{7}{2}$.

On the other hand, let us recall that the type IIB supergravity 
spontaneous compactification on AdS$_5 \times S^5$ for the 10-dimensional dilatino leads to two towers of 5-dimensional Kaluza-Klein dilatino modes, $\lambda^{\pm}_k$, whose 5-dimensional masses are $m_k^+=k+\frac{7}{2}$ and $m_k^-=-k-\frac{3}{2}$, respectively \cite{Kim:1985ez,vanNieuwenhuizen:2012zk,vanNieuwenhuizen:2019lbe}. These spinor spherical harmonics on $S^5$ are labelled by a set of five positive integers $(l_1,l_2,l_3,l_4,l_5)$, which fulfil the relations $l_5 \ge l_4 \ge l_3 \ge l_2 \ge l_1 \ge 0$. Also, notice that there is the identification $l \equiv l_5 \equiv k$. Recall that the second Dynkin integer $r_2$ is now $k$. The degeneracies of the above five-dimensional dilatino modes are given by
\begin{equation}
d_{\text{spinor}}(5, k) = 4 \left[ \left( \begin{matrix} 5+k \\ k  \end{matrix}  \right) - \left( \begin{matrix} 5+k-1 \\ k-1  \end{matrix}  \right) \right] \, , \label{deg-spinor-S5}
\end{equation}
always for $k \ge 0$.

Now, let us work out some relevant examples for us.
Consider first the case $k=0$. The dimension of the $[1, 0, 0]$ representation is given by the ratio between the values of the following Young tableaux: ${\scriptsize\young(4)}$ and ${\scriptsize\young(1)}$, whose values are 4 and 1, respectively. This corresponds to the degeneracy of the mass $m^-_0=-3/2$ given by equation (\ref{deg-spinor-S5}), {\it i.e.} there are four dilatino modes corresponding to the $(0, 0, 0, 0, 0)_a$ spinor spherical harmonics of $S^5$. The sub-index $a=1, \cdots, 4$ labels each of these spinor spherical harmonics.

For $k=1$ we have ${\cal {O}}_{k=1}^{I_1, (6)}(x) = C^{I_1, (6)}_{i_1} \text{Tr}(F_+ \lambda_{{\cal {N}}=4} X_{i_1})(x)$, corresponding to $\tau=4$, and these operators are in the {\bf 20$^*$} irreducible representation of $SU(4)_R$, which is labelled as $[1, 1, 0]$. Its dimension is given by the ratio between the values of the following Young tableaux: 
\begin{equation}
\Yvcentermath1
\Yboxdim{12pt}
{\scriptsize\young(45,3)} \,\,\,\,\, \text{and}  \,\,\,\,\,
{\scriptsize\young(31,1)} \, ,
\end{equation}
whose values are $4 \cdot 5 \cdot 3$ and $3 \cdot 1 \cdot 1$, respectively, thus obtaining 20 as the dimension of this representation. This number is the same as the number of the mass degeneracy of the corresponding five-dimensional dilatino modes given by $(1,1,1,1,1)_a$, $(1,1,1,1,0)_a$, $(1,1,1,0,0)_a$, $(1,1,0,0,0)_a$ and $(1,0,0,0,0)_a$, each of which has four spinors associated.

When $k=2$ the operators are ${\cal {O}}_{k=2}^{I_2, (6)}(x) = C^{I_2, (6)}_{i_1 i_2} \text{Tr}(F_+ \lambda_{{\cal {N}}=4} X_{i_1} X_{i_2})(x)$, which correspond to $\tau=5$, being operators in the {\bf 60$^*$} irreducible representation of $SU(4)_R$. In this case this is the $[1, 2, 0]$ representation. Now, the dimension is given by the ratio between the values of the following Young tableaux: 
\begin{equation}
\Yvcentermath1
\Yboxdim{12pt}
{\scriptsize\young(456,34)} \,\,\,\,\, \text{and}  \,\,\,\,\,
{\scriptsize\young(431,21)} \, ,
\end{equation}
whose values are $4 \cdot 5 \cdot 6 \cdot 3 \cdot 4$ and $4 \cdot 3 \cdot 1 \cdot 2 \cdot 1$, respectively, which gives 60. This number corresponds to the mass degeneracy of the corresponding five-dimensional dilatino modes: 
\begin{eqnarray}
&& (2,2,2,2,2)_a, (2,2,2,2,1)_a, (2,2,2,2,0)_a, (2,2,2,1,1)_a, (2,2,2,1,0)_a, \nonumber \\
&& (2,2,2,0,0)_a, (2,2,1,1,1)_a, (2,2,1,1,0)_a, (2,2,1,0,0)_a, (2,2,0,0,0)_a, \nonumber \\
&& (2,1,1,1,1)_a, (2,1,1,1,0)_a, (2,1,1,0,0)_a, (2,1,0,0,0)_a, (2,0,0,0,0)_a \, . \nonumber
\end{eqnarray}
As before each of them has four spinors associated.

In the next section we will show that the Pauli terms in the five-dimensional supergravity action allow for mixing of Kaluza-Klein dilatino modes which belong to different mass towers. Thus, also local operators of the form ${\cal {O}}_k^{I_k, (13)}(x) = C^{I_k, (13)}_{i_1 \dots i_k} \, \text{Tr}(F_+^2 {\bar{\lambda}}_{{\cal {N}}=4} X_{i_1} \dots X_{i_k})(x)$ give relevant contributions to the structure functions we are interested in. The corresponding irreducible representation of $SU(4)_R$ are $[0, k, 1]$. These operators transform in the $(0, 1/2)$ representation of $SO(1,3)$ and their conformal dimensions are $\Delta=k+11/2$. An important point to keep in mind is that for ${\cal {O}}_k^{I_k, (13)}(x)$ operators the relation between the twist and $k$ is now $\tau=k+5$, which is different from the ${\cal {O}}_k^{I_k, (6)}(x)$ operators. Therefore, for $k=0$ which corresponds to twist-5 spin-1/2 operators ${\cal {O}}_{k=0}^{I_0, (13)}(x)$, there are four of such operators which transform in the ${\bf 4}$ irreducible representation, being this number obtained from the ratio of the values of the Young tableaux: 
\begin{equation}
\Yvcentermath1
\Yboxdim{12pt}
{\scriptsize\young(4,3,2)} \,\,\,\,\, \text{and} \,\,\,\,\,
{\scriptsize\young(3,2,1)} \, .
\end{equation}
Next, let us consider the case $k=1$, then ${\cal {O}}_{k=1}^{I_1, (13)}(x) = C^{I_1, (13)}_{i_1} \text{Tr}(F_+^2 {\bar {\lambda}}_{{\cal {N}}=4} X_{i_1})(x)$, corresponding to $\tau=6$, and these operators transform in the {\bf 20} irreducible representation of $SU(4)_R$, which is labelled as $[0, 1, 1]$. The dimension is given by the ratio between the values of these two Young tableaux:
\begin{equation}
\Yvcentermath1
\Yboxdim{12pt}
{\scriptsize\young(45,34,2)} \,\,\,\,\, \text{and}  \,\,\,\,\,
{\scriptsize\young(42,31,1)} \, .
\end{equation}
When $k=2$, then ${\cal {O}}_{k=2}^{I_2, (13)}(x) = C^{I_2, (13)}_{i_1 i_2} \text{Tr}(F_+^2 {\bar {\lambda}}_{{\cal {N}}=4} X_{i_1}X_{i_2})(x)$, corresponding to $\tau=7$, and these operators are in the {\bf 60} irreducible representation of $SU(4)_R$, which is labelled as $[0, 2, 1]$. The dimension is given by the ratio between the values of the following tableaux:
\begin{equation}
\Yvcentermath1
\Yboxdim{12pt}
{\scriptsize\young(456,345,2)} \,\,\,\,\, \text{and}  \,\,\,\,\,
{\scriptsize\young(532,421,1)} \, .
\end{equation}

Thus, we have discussed the identification of the second Dynkin label $k$ of each irreducible representation of $SU(4)_R$ with the number $l_5$ of the spinor spherical harmonic on $S^5$. Another important point that will be specified later is the relation between $l_1$ and the charge ${\cal {Q}}$ given in equation (\ref{thetachargeeigenstates}).

%
\section{The dual type IIB supergravity calculation of the structure functions}
%

In this section we carry out the holographic dual calculation of  the contributions from the single-trace higher-twist spin-1/2 operators to the structure functions. The holographic dual of the large-$N$ limit of $SU(N)$ ${\cal{N}}=4$ SYM theory is given in terms of type IIB supergravity on AdS$_5 \times S^5$. The metric can be written as
\begin{equation}
ds^2= \frac{dz^2 +  \eta_{\mu\nu} dx^\mu dx^\nu}{z^2}+d\Omega_5^2 \, ,
\end{equation}
where we set to one radius of $S^5$ as well as the scale of the AdS$_5$. The AdS$_5$ indices are $m,n,\cdots= 0,...,4$, the boundary four-dimensional indices are $\mu,\nu,\cdots=0,...,3$, while the $S^5$ indices are $\alpha, \beta, \cdots = 1,...,5$.  The bulk coordinate $z \to 0$ in the UV and we consider a cut-off $z_0=1/\Lambda$ in the IR to induce confinement. This is the so-called hard-wall model. 

The hadronic tensor can be calculated from the matrix elements of two electromagnetic currents inside the hadron by using the optical theorem. Thus, we have to calculate the imaginary part of the tensor $T^{\mu\nu}$ given in equation (\ref{optical-teo}) corresponding to the forward Compton scattering. The Witten's Ansatz allows us to calculate the above matrix elements by evaluating the on-shell supergravity action and taking the sum over all posible intermediate states. Using the covariant type IIB supergravity equations of motion, in \cite{Jorrin:2020cil} we have obtained the effective five-dimensional supergravity action involving two dilatino fields and a massless vector field. We have done it from first principles and therefore we have obtained all the constants from the corresponding angular integrals, which in addition have lead to certain selection rules for the Kaluza-Klein states involved in the fermion interactions. The dilatino field is a right-handed spinor 
\begin{equation}
    \hat{\lambda}(x, y)=\left({\begin{array}{c}
   0\\
   \lambda(x, y) \\
  \end{array} }\right) \, , 
\end{equation}
which can be written as a linear combination of the spinor spherical harmonics on $S^5$ as
\begin{eqnarray}
\lambda(x, y)=\sum_k \left(\lambda^+_k(x) \Theta_k^+(y) + \lambda_k^{-}(x)
\Theta_k^-(y)\right) \, , \label{spinor-expansion}
\end{eqnarray}
where $\Theta_k^+(y)$ and  $\Theta_k^-(y)$ satisfy the Dirac equations on the 5-sphere
\begin{eqnarray}
\tau^{\alpha}D_{\alpha} \Theta_k^{\pm}= \mp i
\left(k+\frac{5}{2} \right)\Theta_k^{\pm} \ \ \ \ \ \ \text{with} \ \
\  k \geq 0 \, . \label{thetakpm}
\end{eqnarray}
Also, the spinor spherical harmonics turn out to be charge eigentates satisfying 
\begin{eqnarray}
\left( v^{\alpha}D_{\alpha} - \frac{1}{4} \tau^\alpha \tau^\gamma \nabla_\gamma v_\alpha \right) \Theta_k^{\pm}= - i {\cal {Q}} \, \Theta_k^{\pm} \, . \label{thetachargeeigenstates}
\end{eqnarray}
$\lambda^{\pm}_k$ are Kaluza-Klein fields with masses given by $m^{\pm}_k$ defined on the AdS$_5$, while the superscripts $\pm$ indicate the two towers of masses associated with the  irreducible representations {\bf 4$^*$}, {\bf 20$^*$}, {\bf 60$^*$}, $\cdots$ ($-$), or {\bf 4}, {\bf 20}, {\bf 60}, $\cdots$ ($+$) of the $SO(6) \sim SU(4)$ isometry group. Coordinates $x$ and $y$ are on AdS$_5$ and on $S^5$, respectively. Gamma matrices in AdS$_5$ and $S^5$ are denoted by $\gamma^m$ and $\tau^{\alpha}$, respectively. They satisfy the Clifford algebra 
\begin{eqnarray}
    \{\gamma_{\hat{a}},\gamma_{\hat{b}} \}=2 \eta_{\hat{a} \hat{b}} \, , \ \ \ \ \{ \tau_{\hat{\alpha}}, \tau_{\hat{\beta}} \}=2 \delta_{\hat{\alpha} \hat{\beta}} \, ,
\end{eqnarray} 
where indices $\hat{a},\hat{b},\hat{c},\dots$ and $\hat{\alpha}, \hat{\beta}, \hat{\gamma}, \dots$ correspond to flat space-time. The vielbein field $e_{a}^{\hat{b}}$ is used to relate the AdS indices to flat-space indices. Analogously, $e_{\alpha}^{\hat{\beta}}$ is associated with the $S^5$.

The structure functions of polarised spin-1/2 hadrons related to operators of the type ${\cal {O}}_k^{I_k, (6)}$ in the ${\cal{N}}=4$ SYM theory can be calculated by using the effective action at leading order obtained in \cite{Jorrin:2020cil}. The following interaction terms have been derived from first principles, {\it i.e.} from direct dimensional reduction on $S^5$ for the dilatino terms at leading order in type IIB supergravity,
\begin{eqnarray}
S_{int}&=& K \int dz \ d^4x \sqrt{-g_{AdS_5}} \times \nonumber \\
&& \ \left( i \frac{{\cal{Q}}}{3}
\bar{\lambda}^-_{k} \gamma^a B^1_a\lambda^-_{k} +i \ \frac{b^{-, -}_{1 k j}}{12} \bar{\lambda}^-_{j} F^{ab}  \Sigma_{ab} \lambda^-_k + i  \frac{b^{+, -}_{1 k j}}{12} \bar{\lambda}^+_{j} F^{ab} \Sigma_{ab} \lambda^-_k\right) \, , \label{five-dimensional-action} 
\end{eqnarray}
where 
\begin{eqnarray}
b^{\pm,-}_{1 k j}&=&\left(1+ 2 \left(k\mp j+\frac{5}{2}\mp\frac{5}{2}\right) \right) \int d \Omega_5  (\Theta^{\pm}_{j})^{\dag} \tau_{\alpha}v^{\alpha}\Theta_{k} +4{\cal{Q}}  \int   d \Omega_5 (\Theta^{\pm}_{j})^{\dag}  \Theta^-_{k}  \, .
\end{eqnarray}
$K$ is a normalization constant that can be calculated by comparison with the type IIB supergravity action of reference \cite{DallAgata:1998ahf}. In addition, $B_a^1$ is the massless Maxwell-Einstein field in AdS$_5$ defined in equation (\ref{Bfield}), while $F_{ab}=\nabla_a B_b^1-\nabla_b B_a^1$ and $\Sigma_{ab}=\frac{1}{4}(\gamma_a \gamma_b-\gamma_b \gamma_a)$. Early references for the covariant equations of motions of type IIB supergravity fields are \cite{Schwarz:1983wa,Howe:1983sra,Schwarz:1983qr}.

The first term in the action (\ref{five-dimensional-action}) corresponds to the minimal-coupling interaction used to calculate the structure functions of a DIS process in \cite{Polchinski:2002jw,Gao:2009ze} and the vector-spinor-spinor three-point function in reference \cite{Mueck:1998iz}. This coupling only connects states in the same irreducible representation (which have the same twist). The other interactions are Pauli terms whose strengths are given by the coefficients $b_{1kj}$ calculated from the angular integrals of spinor spherical harmonics. We can separate contributions having states in the same irreducible representation, and mixing of states from different irreducible representations of $SU(4)$, which are given by the second and the third terms, respectively.   

Before studying the selection rules for higher-twist operators, we briefly review the solutions in AdS$_5$ of non-normalizable modes of the vector field and the normalizable modes of dilatini, which correspond to the holographic dual fields of the electromagnetic current and the hadronic states, respectively. These solutions will be inserted in the effective five-dimensional action (\ref{five-dimensional-action}) to calculate the matrix elements of the electromagnetic currents.

The massless vector fields $B^1_{a}$ come from a certain linear combination of off-diagonal fluctuations of the metric tensor
and vector fluctuations of the Ramond-Ramond four-field potential in the following way,
\begin{equation}
B_a^1(x) \equiv A_a^1(x) - 16 \Phi_a^1(x) \, , \label{Bfield}
\end{equation}
where $A_a^1(x)$ is defined by the metric fluctuation as
\begin{equation}
h_{a \alpha} = \sum_{I_5} A_a^{I_5}(x) \, Y^{I_5}_\alpha(y) \, ,
\end{equation}
and $\Phi_a^1(x)$ is given in terms of the mode expansion of the Ramond-Ramond field as
\begin{equation}
a_{a \alpha\beta\gamma} = \sum_{I_5} \Phi_a^{I_5}(x)\, \epsilon_{\alpha\beta\gamma\delta\epsilon} \, \nabla^\delta Y^{I_5 \epsilon}(y) \, .
\end{equation}
In particular, the index $I_5$ denotes the set of numbers $(l_5,l_4,l_3,l_2,l_1)$ for the vector spherical harmonics on $S^5$, $Y^{I_5 \epsilon}(y)$. The corresponding masses of these vector fields are given by $M^2_{B, l}=l^2-1$ with $l \ge 1$, therefore they only depend on $l \equiv l_5$ and, in terms of the irreducible representations of $SU(4)$ they transform in the {\bf 15}, {\bf 64}, {\bf 175}, $\cdots$, for $l=1$, 2, 3, $\cdots$. For the holographic DIS calculation we only need to consider the massless vector fields, {\it i.e.} $B_a^1(x)$, which are the 15 Yang-Mills fields of $SU(4) \sim SO(6)$. In addition, in this case the vector spherical harmonics are Killing vectors of $S^5$. The gauge fields satisfy the following Einstein-Maxwell equation of motions in AdS$_5$
\begin{eqnarray}
    \nabla_a F^{ab} &=& 0\, , \label{B1} \\
\partial^{\mu}B^1_{\mu}+z \partial_z\left( \frac{B^1_z}{z} \right)&=&0 \, . \label{B2}
\end{eqnarray}
The second equation gives a Lorentz-type gauge fixing condition. Then, the non-normalizable modes which are dual to the hadronic current on the boundary satisfy the following boundary condition
\begin{equation}
B^1_{\mu}(x_\nu,z\to0)= n_{\mu} \, e^{i q \cdot x} \, . \label{Bbc}
\end{equation}
Thus, the solutions of the equations (\ref{B1}) and (\ref{B2}) with the boundary condition (\ref{Bbc}) read
\begin{eqnarray}
B^1_{\mu}(x_\nu, z)=n_{\mu} \, e^{i q \cdot x}  \, q \, z \, K_1(q z), \ \ \ \ \ \  B^1_z(x_\nu, z)=i \, n \cdot q \, e^{i q \cdot x} \, z \,  K_{0}(q z) \, .
\end{eqnarray}
On the other hand, the dilatini satisfy the Dirac equation in AdS$_5$ with the hard-wall boundary condition at the IR, needed in order to break the conformal symmetry and induce color confinement. Thus, we impose Dirichlet boundary conditions at the IR cut-off $z_0=1/\Lambda$. In addition, in the ultraviolet region ($z\to 0$) the boundary condition is fixed by choosing the normalizable mode for the initial and final hadronic states.
The Dirac equation in AdS$_5$ reads
\begin{eqnarray}
\left(z \gamma^{m} \partial_{m}-2 \gamma^5-k-\frac{3}{2} \right)\lambda^-_k=0 \, , \label{dirac-dif}
\end{eqnarray}
being the normalizable solution 
\begin{equation}
\lambda_k^-(x_\nu,z) = C e^{i P \cdot x}z^{\frac{5}{2}}\left(  J_{\tau-2}(M z) P_{+} +  J_{\tau-1}(M z)P_{-} \right)u_{\sigma} \, , \label{spinor-pos}
\end{equation} 
where the projectors are
\begin{equation}
P_{\pm}=\frac{(I\pm\gamma^5)}{2}  \, ,
\end{equation}
while $P^\mu$ is the four-momentum of the hadron, and the solution has been expressed in terms of Bessel functions of the first kind and four-dimensional Dirac spinors $u_{\sigma}$. These spinors satisfy $\gamma^{\mu} P_{\mu} u_{\sigma}= i M u_{\sigma}$ with $P^2=-M^2$. The twist $\tau= \Delta-\frac{1}{2}=m_k + 3/2$ corresponds to the ${\cal {N}}=4$ SYM  operator ${\cal {O}}_k^{I_k, (6)}$. The constant $C=c_{\cal {X}}' M_{\cal {X}}^{1/2}/z_0^{1/2}$ can be expressed in terms of another dimensionless constant $c_{\cal {X}}'$, following the normalization used in reference \cite{Polchinski:2002jw}. The identity matrix is indicated as $I$. In addition, the bulk solutions for the fields $\lambda^+$ dual to ${\cal{O}}_k^{I_k, (13)}$ operators are calculated in the same way.

The coefficients in the effective action (\ref{five-dimensional-action}) are given in terms of  integrals of spinor spherical harmonics on $S^5$, and they lead to the selection rules for the intermediate states in the forward Compton scattering. Recall that the spinor spherical harmonics have five quantum numbers $(l_5,l_4, l_3,l_2,l_1)_a$, which satisfy the conditions $l_5 \geq l_4 \geq l_3 \geq l_2 \geq l_1 \geq 0$.  We also use the subscripts $a=1, 2, 3, 4$.  In particular, $l_5=k$ is related to the twist, while the $l_1$ index is associated with the charge ${\cal{Q}}=\pm (l_1+\frac{1}{2})$. For $k=0$, from the dual SYM theory point of view, the operators ${\cal{O}}^{I_0, (6)}_{k=0}$  belong to the $\textbf{4*}$ irreducible representation of $SU(4)_R$. Thus, there are four of these type of operators, and we have explicitly verified that the final result for the structure functions is the same for all these operators belonging to the $\textbf{4*}$ representation. Similarly, for $k=1$ there are 20 Kaluza-Klein states (and operators) which can be separated in 5 sets, leading to the same structure functions within each set. An analogous situation occurs for $k=2$ where from the 60 Kaluza-Klein states (and operators) there are 15 different sets with the same structure functions for the four states within the same set. We have checked these results.

The minimal coupling only connects states with the same quantum numbers $(l_5,l_4, l_3,l_2,l_1)_a$ on $S^5$ and belong to the same irreducible representation of $SU(4) \sim SO(6)$. The contributions to the structure functions associated with the minimal coupling are denoted by $F_i^m$ and $g_i^m$. They have been calculated in \cite{Jorrin:2020cil} and their explicit dependence in $q$ and $x$ are detailed in the Appendix \ref{Appendix-A}. The other terms of the effective action (\ref{five-dimensional-action}) are Pauli interactions with coefficients $b_{1kj}$. In the second term we can indentify the interaction between the gauge field and two dilatini of the same Kaluza-Klein mass tower. Therefore, they correspond to operators in the same irreducible representation of $SU(4)_R$. The angular integrals only connect states with equal twist. The matrix elements can be calculated by solving the $z$-integrals on AdS$_5$. In Appendix \ref{Appendix-A} these functions are also written in detail, and we denote them by a superscript $P$, namely: $F_i^P$ and $g_i^P$. These interactions can be separated into two sets. The first set is constituted by diagrams which have intermediate states with the same quantum numbers $(l_5,l_4, l_3,l_2,l_1)_a$ as the incident dilatino. Thus, we can calculate their coefficients by the following integral 
\begin{eqnarray}
b^{-,-}_{1 k k}&=&\left(11+4k\right)\int d \Omega_5  (\Theta^{-}_{k}){\dag} \tau_{\alpha}v^{\alpha}\Theta_{k}^- +4 {\cal{Q}}  \, .
\end{eqnarray}
We define the constant
\begin{eqnarray}
\beta_P=\frac{b_{1 k k}^{-,-}}{12} \, .
\end{eqnarray}
This Pauli interaction term is particularly important since when we calculate the product of the one-point function of the electromagnetic current and its complex conjugate in equation (\ref{optical-teo}), there is a cross contribution corresponding to a Feynman-Witten diagram which also includes the minimal coupling. This leads to matrix elements of the hadronic tensor of the form
\begin{equation}
    n_{\mu} n_{\nu}   W^{c\mu \nu}=n_{\mu} n_{\nu} \ 2 \pi^2 \sum_{{\cal {X}}}\delta(M_{\cal {X}}^2+(P+q)^2) (\langle J_m^{\mu}(0)\rangle \langle J_P^{*\nu}(0)\rangle+\langle J_P^{\mu}(0)\rangle \langle J_m^{*\nu}(0)\rangle) \, .
\end{equation}
The structure functions from cross terms are indicated with the superscript $c$: $F^c_i$ and $g_i^c$ and they are explicitly shown in Appendix \ref{Appendix-A}.

In figure \ref{fig-sk-DIS} we illustrate the Feynman-Witten diagrams needed to calculated the matrix elements of the electromagnetic currents inside the hadron. The first matrix elements correspond to the same incoming and outgoing state, with the minimal coupling and the Pauli term (dotted vertex) discussed above. On the other hand, as already mentioned the Pauli diagrams with final states belonging to same irreducible representations of $SU(4)$ ({\it i.e.} equal $k=l_5$), but different $l_4, l_3,l_2,l_1$ numbers, give the same contributions to the structure functions. However, they do not lead to cross terms involving the minimal coupling. Now, we define the following constant     
\begin{eqnarray}
\beta^2_{Pm}&=&\sum_{I_5} \left(\frac{b^{-,-}_{1 k k}}{12}\right)^2=\sum_{I_5} \left(\frac{\left(11+4k\right) \int d \Omega_5  (\Theta^{+}_{k})^{\dag} \tau_{\alpha}v^{\alpha}\Theta_{k}^-}{12} \, \right)^2 \, ,
\end{eqnarray}
where $I_5$ indicates the quantum numbers $(l_5,l_4, l_3,l_2,l_1)_a$ of the possible intermediate states.

The last term of equation (\ref{five-dimensional-action}) couples fermionic modes of different Kaluza-Klein towers of type IIB supergravity compactified on $S^5$. Considering Feynman-Witten diagrams with incoming states dual to the operators ${\cal{O}}_k^{I_k, (6)}$, the intermediate states will be dual to ${\cal{O}}_{k'}^{I_{k'}, (13)}$ with $k'=k+1$ or $k'=k-1$. The selection rule $k'=k \pm 1$ is obtained from the angular integral of the spinor spherical harmonics\footnote{These selection rules relate the conformal dimensions of the incident and intermediate states through $\Delta_{\cal {X}}=\Delta_i+2\pm1$.}, being the matrix elements calculated using the Feynman-Witten diagrams of the second and third lines in figure \ref{fig-sk-DIS}.

\begin{figure}[H]
\centering
\includegraphics[scale=0.4]{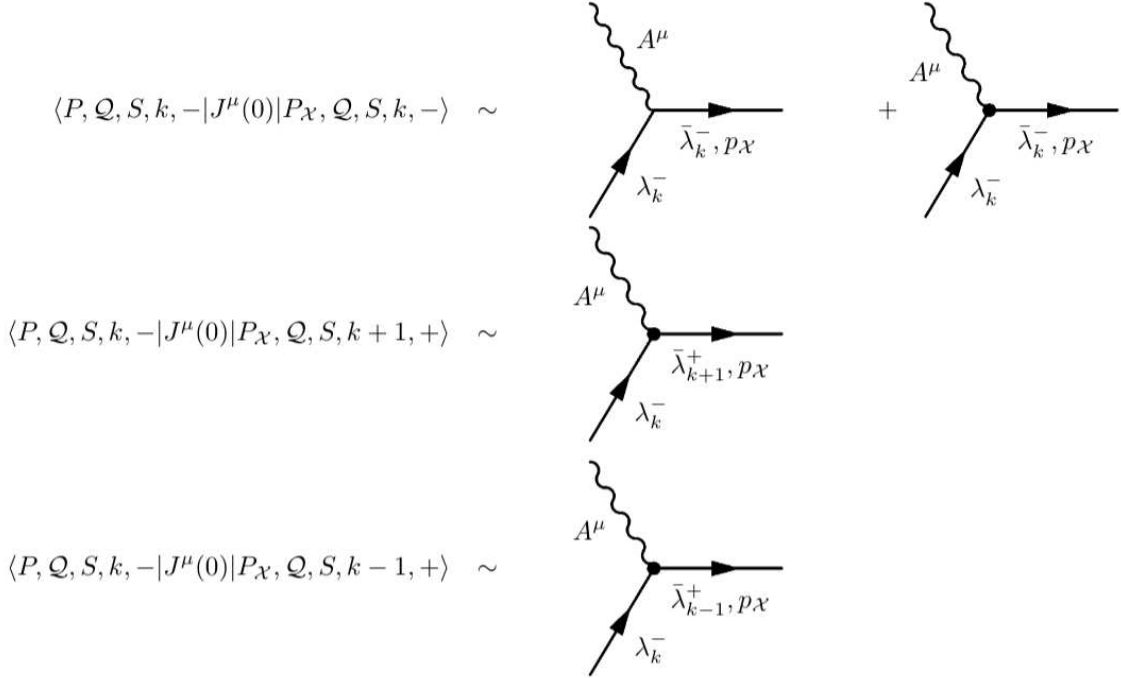}
\caption{{\small Matrix elements of the electromagnetic current inside the hadron are written on the left. The minimal coupling is represented by the first diagram. The diagrams with dotted vertices correspond to different Pauli terms. Selection rules allow for these three Pauli diagrams.}} 
\label{fig-sk-DIS}
\end{figure}

The constant $\beta^2_-$ for the case $k'=k-1$ is calculated in terms of the coefficients $b_{1kj}^{+,-}$ of the effective action as follows
\begin{eqnarray}
\beta_-^2&=&\sum_{I_5} \left(\frac{b^{+,-}_{1 \ k \ k-1}}{12}\right)^2=\sum_{I_5} \left(\frac{\int d \Omega_5  (\Theta^{+}_{k-1})^{\dag} \tau_{\alpha}v^{\alpha}\Theta_{k}^-}{4} \, \right)^2 \, .
\end{eqnarray}
The corresponding contributions to the structure functions are given by 
\begin{eqnarray}
F^{P-}_1&=&\frac{F_3^{P-}}{2}=g_1^{P-}=\frac{g_5^{P-}}{2}= \frac{1}{2} |a_0|^2 \left(\frac{\Lambda^2}{q^2}\right)^{\tau-1}(1-x)^{\tau-2} x^{\tau+1} \Gamma(\tau)^2 \, , \\
%
%
%
F^{P-}_2&=&g_4^{P-} =|a_0|^2 \left(\frac{\Lambda^2}{q^2}\right)^{\tau-1}(1-x)^{\tau-3} x^{\tau+1} \Gamma(\tau)^2(1+x(3+\tau(-2+x(\tau-2)))) \,  ,  \\
%
%
%
g_2^{P-}&=& -\frac{1}{4} |a_0|^2 \left(\frac{\Lambda^2}{q^2}\right)^{\tau-1}\frac{\Gamma(\tau)^2}{\tau-1} x^{\tau+1} (1-x)^{\tau-3} \left[- \tau-1+x\left(x^{5}(\tau+1)\tau (\tau-2)+(1+\tau)^2 \right.\right. \nonumber \\
&& \left.\left. -x^4(\tau-2)\tau (7+3\tau)+6x^2(2+(3-2\tau)\tau) \right. \right. \nonumber\\
&&  \left. \left. -x (12+(\tau-5)\tau)+x^3(\tau-3)(2+3\tau(5+\tau)) \right) \right] \, , \\
%
%
%
g_3^{P-}&=& |a_0|^2 \left(\frac{\Lambda^2}{q^2}\right)^{\tau-1} \frac{\Gamma(\tau)^2}{\tau-1} x^{\tau+1}(1-x)^{\tau-3} \times \nonumber \\
&&\left( 3-\tau+x\left( -15-4(x-3)x-2\tau+x(33-2x(18+(x-8)x))\tau   \right. \right. \nonumber \\
&&  \left. \left.+(1+x(-1+(x-6)x(3+x(x-3))))\tau^2+x^3 (3+x(x-3)\tau^3)  \right) \right) \, ,
\end{eqnarray}
where the constant $a_0=2 \pi c'_i c'_{\cal {X}}2^\tau K$ is written in terms of the factors $c'_i$ and  $c'_{\cal {X}}$ corresponding to the normalization constants of the incident and the intermediate hadronic wave-functions, respectively.
We would like to emphasize that this interaction is not allowed for the case of $\tau=3$ studied in \cite{Jorrin:2020cil}. For this reason it is interesting study its role for higher-twist operators. Finally, we consider the case $k'=k+1$, with a constant given by
\begin{eqnarray}
\beta_+^2&=&\sum_{I_5} \left(\frac{b^{+,-}_{1 \ k \ k+1}}{12}\right)^2=\sum_{I_5} \left(-\frac{\int d \Omega_5  (\Theta^{+}_{k+1})^{\dag} \tau_{\alpha}v^{\alpha}\Theta_{k}^-}{12} \, \right)^2 \, .
\end{eqnarray}
The structure functions associated with these interactions have different dependence on the Bjorken parameter, 
\begin{eqnarray}
F^{P+}_1&=&\frac{F_3^{P+}}{2}=g_1^{P+}=\frac{g_5^{P+}}{2} = \frac{1}{2} |a_0|^2 \left(\frac{\Lambda^2}{q^2}\right)^{\tau-1} x^{-\tau-3}(1-x)^{\tau+2} \Gamma(1+\tau)^2 \times \nonumber \\
&& \left[x^{\tau+1}-(1+\tau)\phantom F _2 F_1\left(\tau+2,\tau+2,\tau+3,\frac{x-1}{x}\right) \right]^2 \, , \\
F^{P+}_2&=&g_4^{P+} = |a_0|^2 \left(\frac{\Lambda^2}{q^2}\right)^{\tau-1} x^{-\tau}(1-x)^{\tau+1} \Gamma(1+\tau)^2 \times \nonumber \\
&&\left[x^{2\tau}+ \frac{1-x}{x^3}(1+\tau)^2 \phantom F _2 F_1\left(\tau+2,\tau+2,\tau+3,\frac{x-1}{x} \right)^2 \right] \, , 
\end{eqnarray}
\begin{eqnarray}
g_2^{P+} &=&  - \frac{1}{4} |a_0|^2 \left(\frac{\Lambda^2}{q^2}\right)^{\tau-1} (1-x)^{\tau} x^{\tau} \Gamma(1+\tau)^2 \times \nonumber \\
&&\left[ 2\frac{1-x}{1-\tau}+ (1-x)^{-2\tau} (1+\tau)^2 (2+\tau) \ B\left(\frac{x-1}{x},\tau+2,-(\tau+1)\right) \right. \nonumber \\
&& \left. \left( \frac{(x-1)^ {\tau}}{1-\tau}+\frac{x(\tau+2) \ B\left(\frac{x-1}{x},\tau+2,-(\tau+1)\right)}{(1-x)^2} \right)\right] \, , \\
g_3^{P+}&=& |a_0|^2 \left(\frac{\Lambda^2}{q^2}\right)^{\tau-1} \frac{1}{\tau-1} 8 (1-x)^{\tau+1} x^{-\tau-3} \Gamma(1+\tau)^2 \times \nonumber \\
&& \left[ 2x^{2\tau+3}+ (1+\tau)^2(x-1) \phantom F _2 F_1\left(\tau+2,\tau+2,\tau+3,\frac{x-1}{x}\right) \right. \nonumber \\
&&  \left. \left( -x^{\tau+1}+ (\tau-1) \phantom F _2 F_1\left(\tau+2,\tau+2,\tau+3,\frac{x-1}{x}\right)\right) \right] \, ,
\end{eqnarray}
which are expressed in terms of the incomplete Beta function, $B(x,a,b)$, and the Hypergeometric function,  $_2F(a,b,c,x)$.

The general form of the structure functions which contain all the contributions for higher-twist operators is given by
\begin{equation}
F_i = \beta^2_{m}F_i^m+\beta^2_P F^P_{i}+ \beta_m \beta_P F^{c}_{i}+\beta^2_{Pm} F^{P}_{i} +\beta^2_+ F^{P+}_i+\beta^2_-F^{P-}_i,
    \label{complete-structure-function}
\end{equation}
and we have a similar expression for the $g_i$ structure functions. For the complete structure functions, {\it i.e.} by adding all the contributions from all allowed interactions from action (\ref{five-dimensional-action}), for any twist, there are the following relations:
\begin{eqnarray}
F_1=\frac{F_3}{2}=g_1=\frac{g_5}{2} \, , \,\,\,\,\,  {\text{and}} \,\,\,\,\,\,\,\, F_2 = g_4 \, .
\end{eqnarray}
Thus, in what follows we will show explicitly $F_1$, $F_2$, $g_2$ and $g_3$, which are independent.

For $\tau=3$ analysed in \cite{Jorrin:2020cil}, the constants  $\beta^2_{-}$ and $\beta^2_{Pm}$ vanish since the incident hadron has $k=l_5=0$ and, consequently all the remaining $l_i$'s are zero. However, for higher-twist operators these contributions are non-zero. In the following subsection we will analyse how each interaction contributes to the structure functions for $\tau=4$ and $\tau=5$.    

~

%
\subsection{The case of twist-4  ${\cal {O}}_{k=1}^{I_1, (6)}=C^{I_1, (6)}_{i_1} \text{Tr}(F_+ \lambda_{{\cal {N}}=4} X_{i_1})$ operators}
%

In this case the incident hadrons correspond to operators of the form ${\cal{O}}^{I_1, (6)}_{k=1}$  which belong to the {\bf 20$^*$} irreducible representation of $SU(4)_R$. They have twist $\tau=4$. The holographic dual fields are represented by dilatino modes with $l_5=k=1$ and their respective quantum numbers are $I_5=(1,l_4,l_3,l_2,l_1)_a$. 

There are 20 independent spinor spherical harmonics with the same Kaluza-Klein mass. This degeneration comes from the possibility of having $I_5=(1,l_4,l_3,l_2,l_1)_a$ values satisfying $l_5=1\geq l_4 \geq l_3 \geq l_2 \geq l_1 \geq 0$ (in fact there are five combinations) times the four degrees of freedom of each spinor, which are parametrized by the subscripts $a=1,...,4$.  The final result only depends on the choice of the $l_i$'s, thus we can separate the 20 initial possibilities in five sets as commented before. Once we choose a certain set of $l_i$'s, the four possible states lead to the same structure functions. Therefore, without loss of generality we choose $a=1$ for the incident hadron. For instance, the spinor spherical harmonic normalized with $I_5=(1,0,0,0,0)_{a=1}$ is
\begin{equation}
    \Theta^{-}_{(1,0,0,0,0)_{a=1}}= 
\frac{e^{-i{\cal{Q}} \theta_1}}{\sqrt{5}\pi^{3/2}} 
\begin{bmatrix} e^{-i \frac{1}{2}( \theta_3 -\theta_5)} 
\cos (\frac{\theta_2}{2}) \cos (\frac{\theta_4}{2})\left(5 i \cos(\theta_5) - \sin(\theta_5) \right) \\ 
-e^{i\frac{1}{2}( \theta_3 +\theta_5)}\sin (\frac{\theta_2}{2}) 
\cos (\frac{\theta_4}{2}) \left(5 i \cos(\theta_5) - \sin(\theta_5) \right) \\ 
-e^{-i\frac{1}{2}( \theta_3 
+\theta_5)}\cos (\frac{\theta_2}{2}) \sin (\frac{\theta_4}{2}) \left(5 i \cos(\theta_5) + \sin(\theta_5) \right)\\
e^{-i\frac{1}{2}( -\theta_3 +\theta_5)} \sin (\frac{\theta_2}{2}) 
\sin (\frac{\theta_4}{2}) \left(5 i \cos(\theta_5) + \sin(\theta_5) \right) \end{bmatrix} \, . \label{spinor10000}
\end{equation}
As mentioned, the angular integrals of the spinor spherical harmonics allow us to obtain the selections rules and the relative coefficients among the contributions given by different terms in the action (\ref{five-dimensional-action}). There is only one state with $l_1=1$, {\it i.e.} with ${\cal{Q}}=\pm \frac{3}{2}$, while the other four have $l_1=0$, {\it i.e.} with ${\cal{Q}}=\pm \frac{1}{2}$. This is important because the charge is a conserved quantity and the integrals are equal to zero if we mix states with different $l_1$. Since we have considered an incident state with $a=1$, the only possibility is the coupling with states with $a=1$ and 3, because they have the same charge. Spinors with $a=2,4$ have charge with a different sign. Tables collecting the details of intermediate states and their coefficients obtained from the angular integrals of the spinor spherical harmonics are displayed in Appendix \ref{Appendix_B}. There we note that the selection rules for $l_i$, with $i=2,3,4$, are given by $l'_i=l_i\pm1$ as long as they satisfy $l_5 \geq l_4 \geq l_3 \geq l_2 \geq l_1$.

\begin{figure}[H]
        \centering
        \begin{subfigure}[b]{0.32\textwidth}
            \centering
            \includegraphics[width=\textwidth]{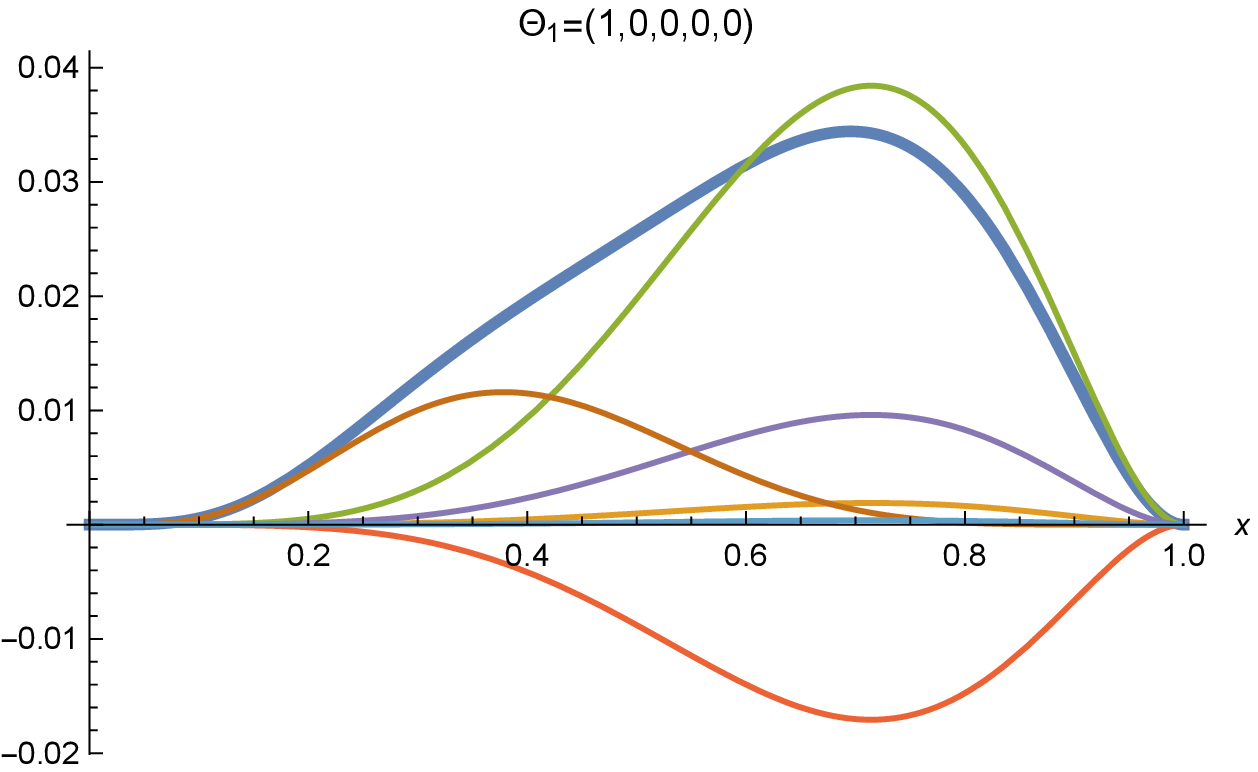}
            \caption[Network2]%
            {{\small $F_1$ }}    
        \end{subfigure}
        \hfill
        \begin{subfigure}[b]{0.32\textwidth} 
            \centering 
            \includegraphics[width=\textwidth]{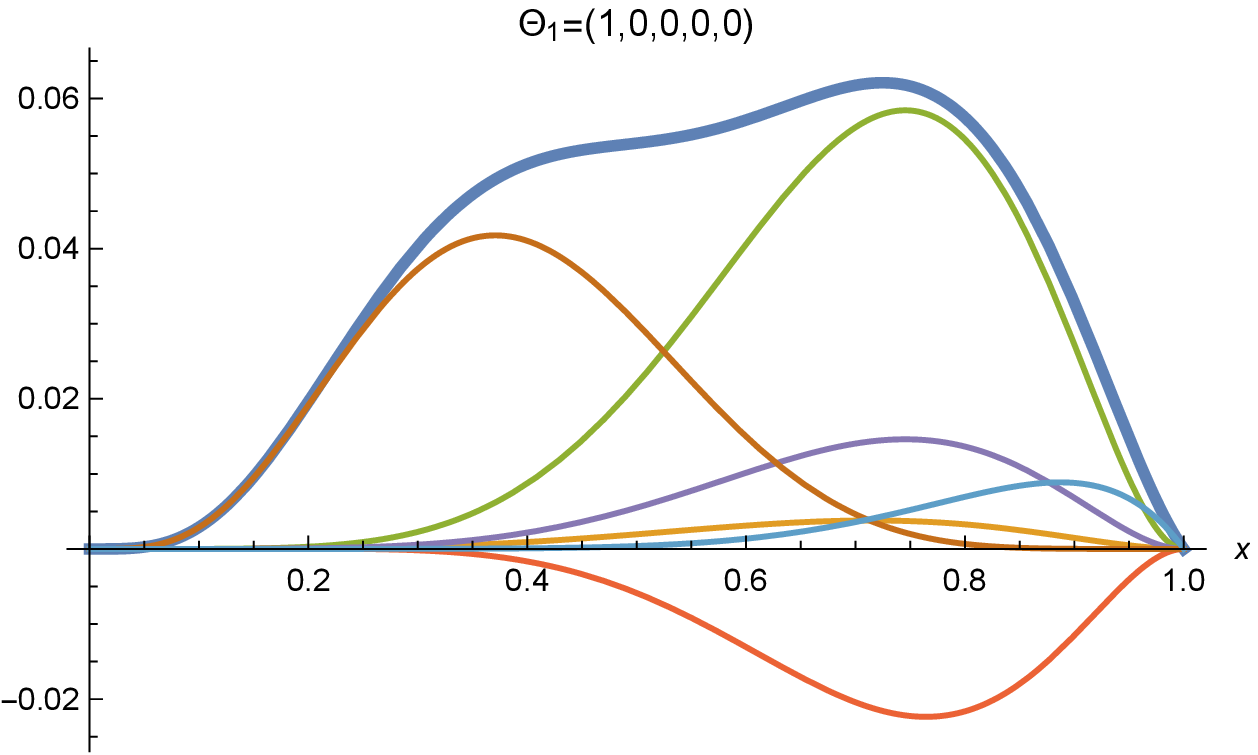}
            \caption[]%
            {{\small $F_2$}}    
        \end{subfigure}
            \hfill
        \begin{subfigure}[b]{0.32\textwidth} 
            \centering 
            \includegraphics[width=\textwidth]{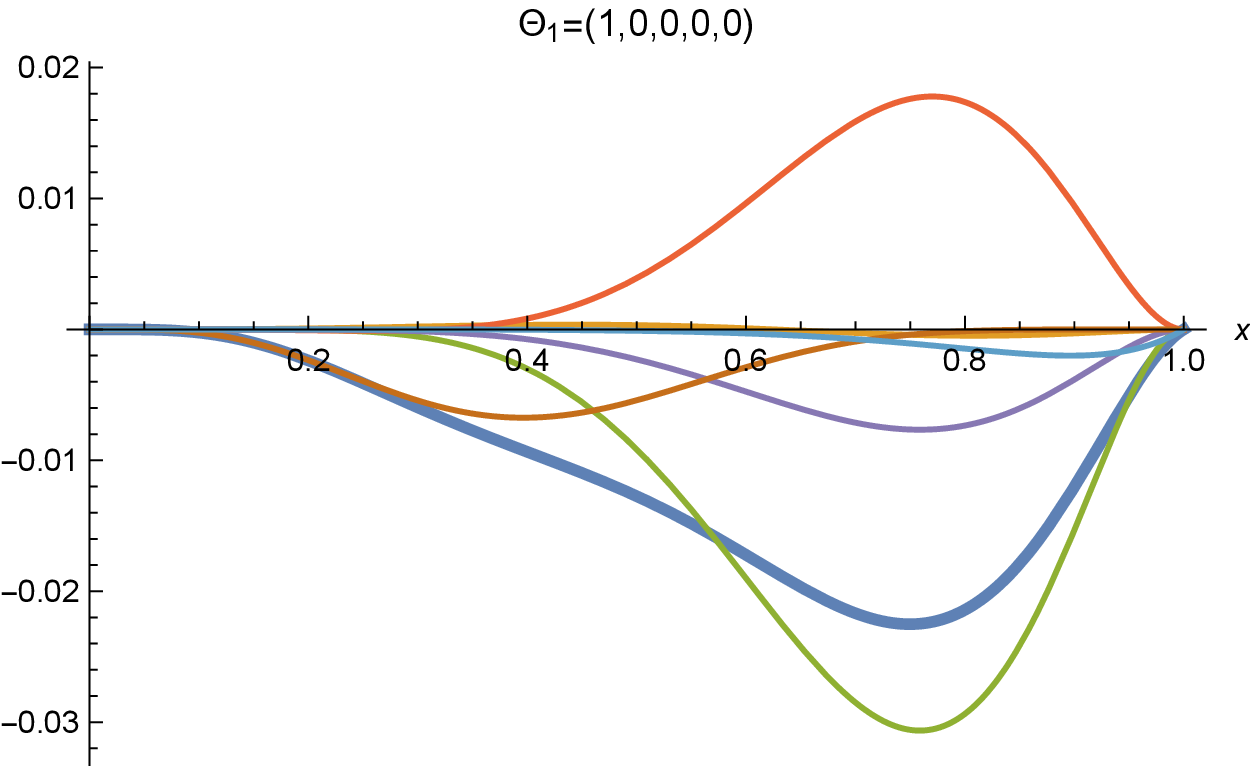}
            \caption[]%
            {{\small $g_2$}}    
        \end{subfigure}
        
        \vskip\baselineskip
        \begin{subfigure}[b]{0.32\textwidth}
            \centering 
            \includegraphics[width=\textwidth]{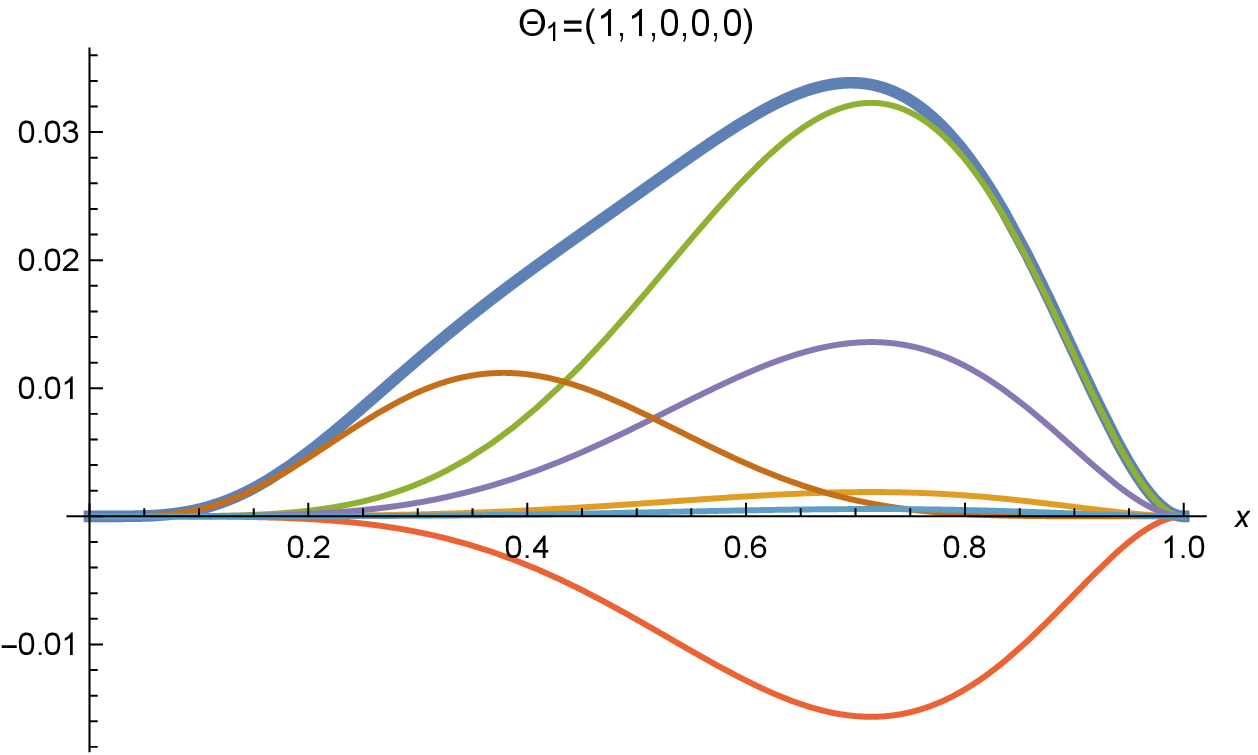}
            \caption[]%
            {{\small $F_1$}}    
        \end{subfigure}
        \hfill
        \begin{subfigure}[b]{0.32\textwidth} 
            \centering 
            \includegraphics[width=\textwidth]{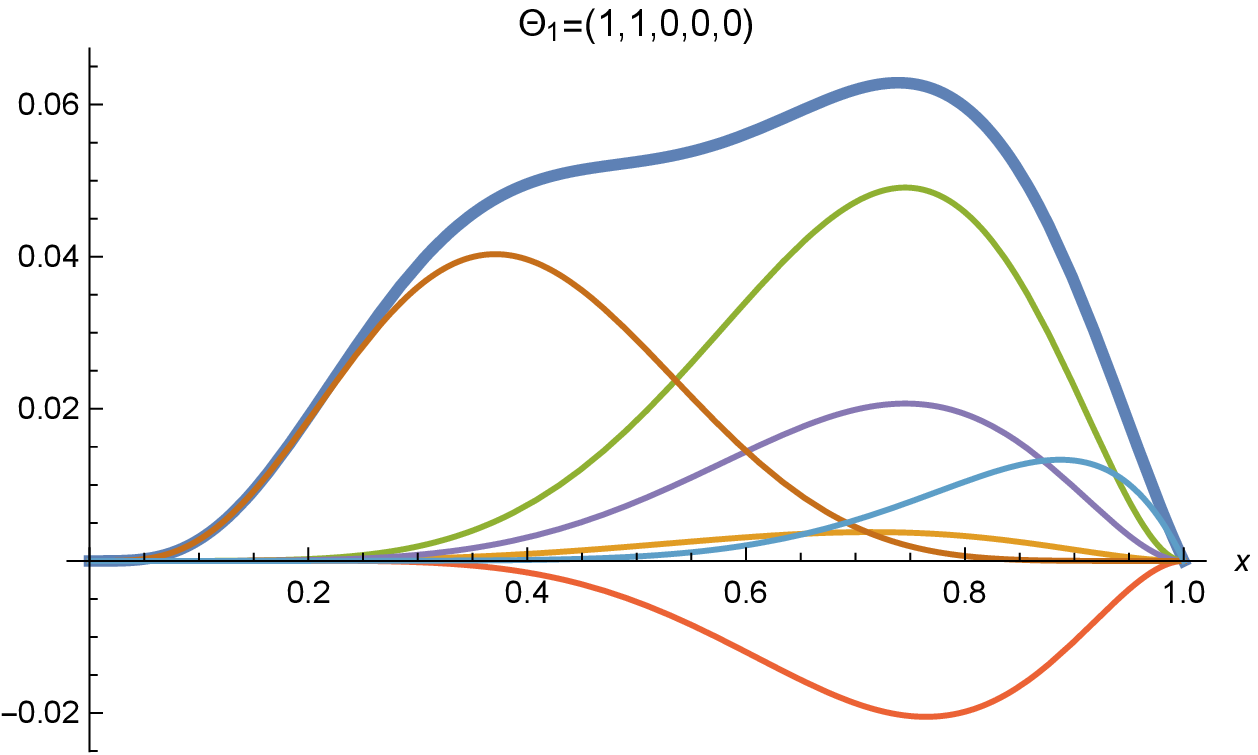}
            \caption[]%
            {{\small $F_2$}}    
        \end{subfigure}
            \hfill
        \begin{subfigure}[b]{0.32\textwidth} 
            \centering 
            \includegraphics[width=\textwidth]{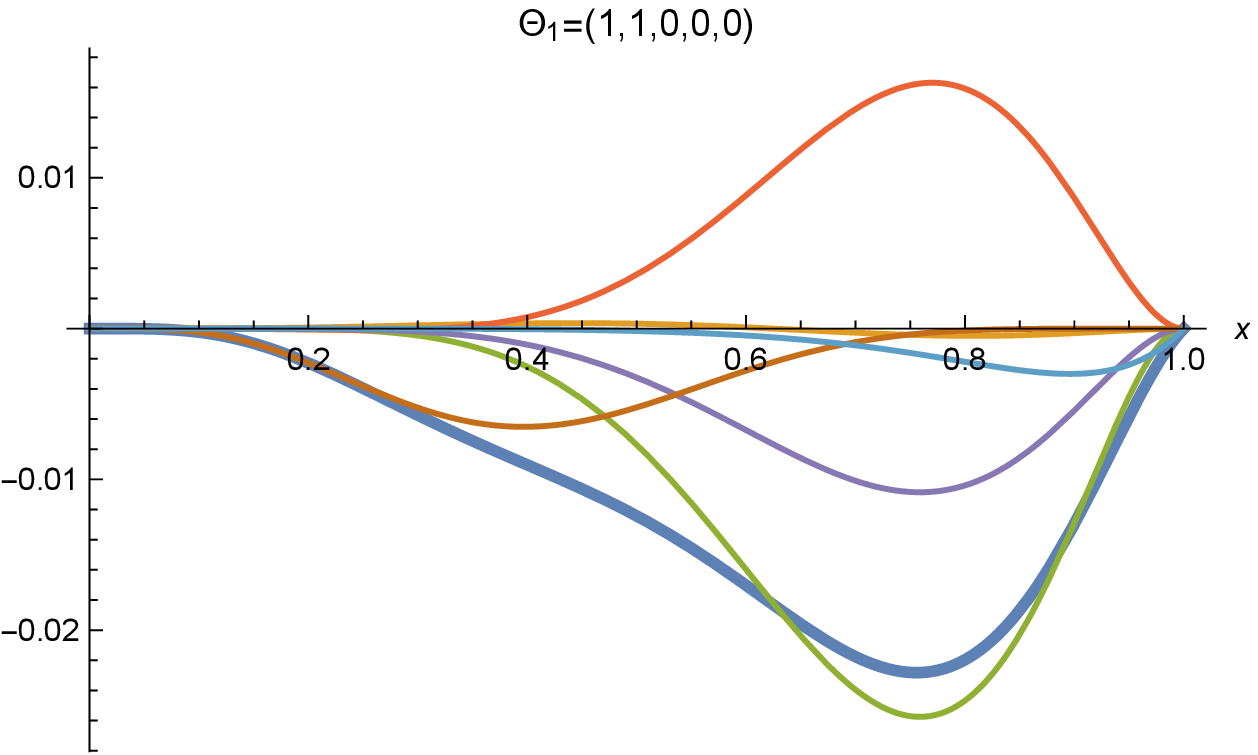}
            \caption[]%
            {{\small $g_2$}}    
        \end{subfigure}
                \vskip\baselineskip
        \begin{subfigure}[b]{0.32\textwidth}
            \centering 
            \includegraphics[width=\textwidth]{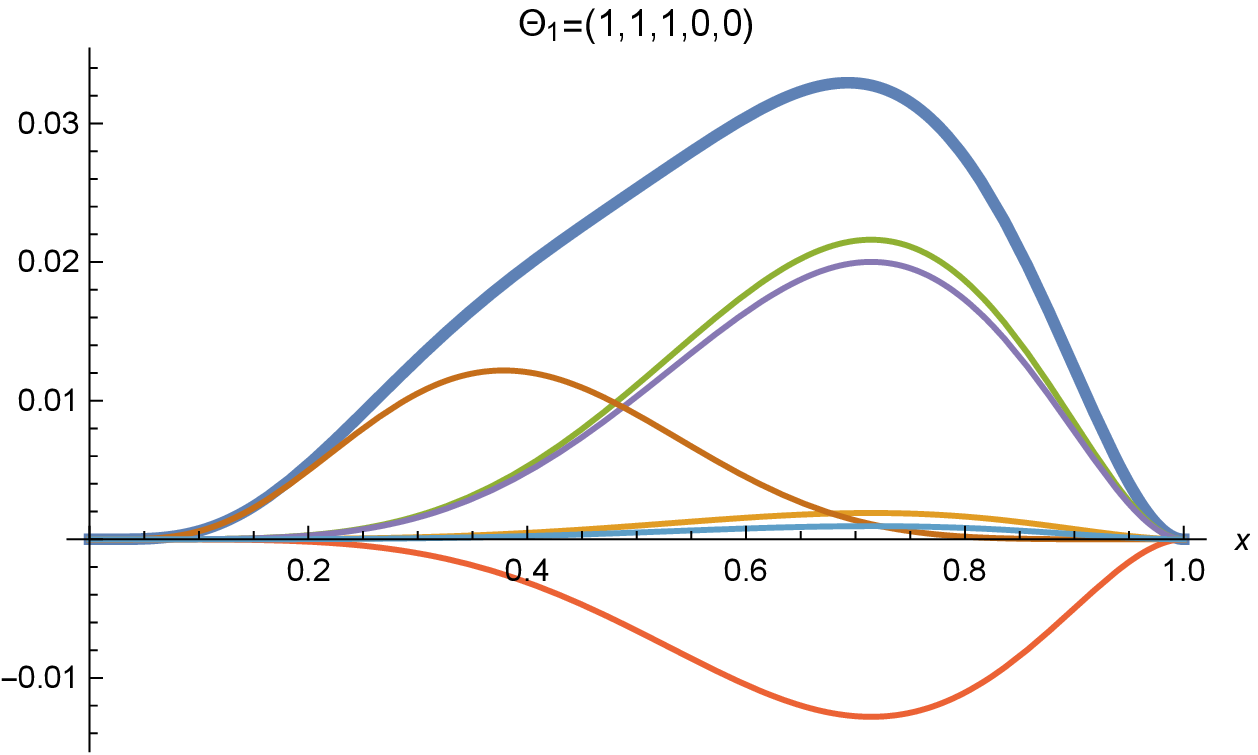}
            \caption[]%
            {{\small $F_1$}}    
        \end{subfigure}
        \hfill
        \begin{subfigure}[b]{0.32\textwidth}   
            \centering 
            \includegraphics[width=\textwidth]{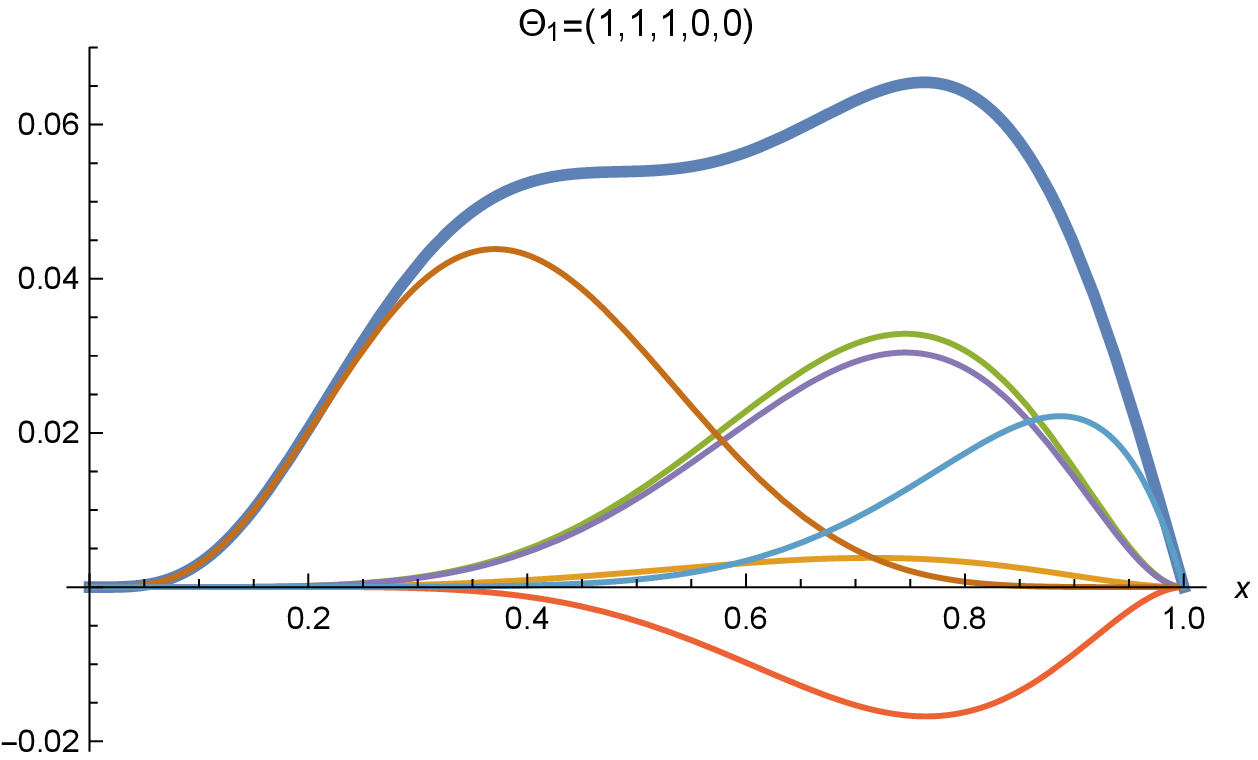}
            \caption[]%
            {{\small $F_2$}}    
        \end{subfigure}
            \hfill
        \begin{subfigure}[b]{0.32\textwidth} 
            \centering 
            \includegraphics[width=\textwidth]{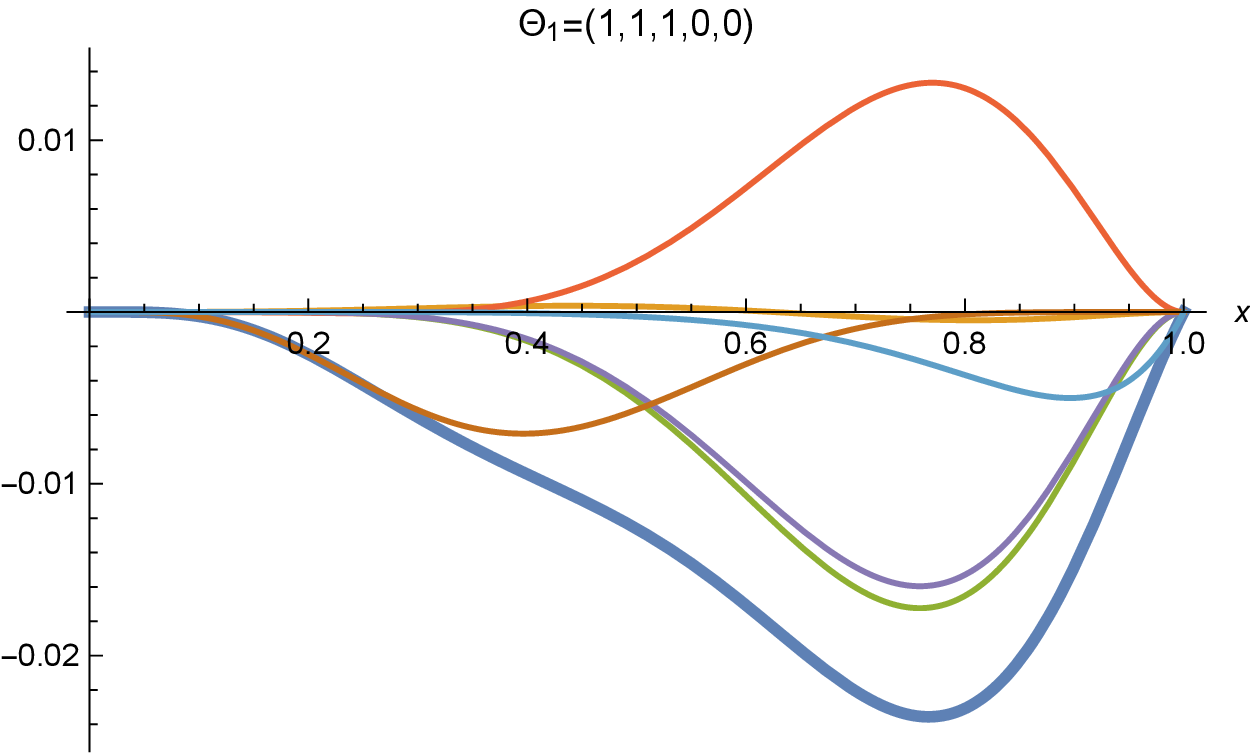}
            \caption[]%
            {{\small $g_2$}}    
        \end{subfigure}
                        \vskip\baselineskip
        \begin{subfigure}[b]{0.32\textwidth} 
            \centering 
            \includegraphics[width=\textwidth]{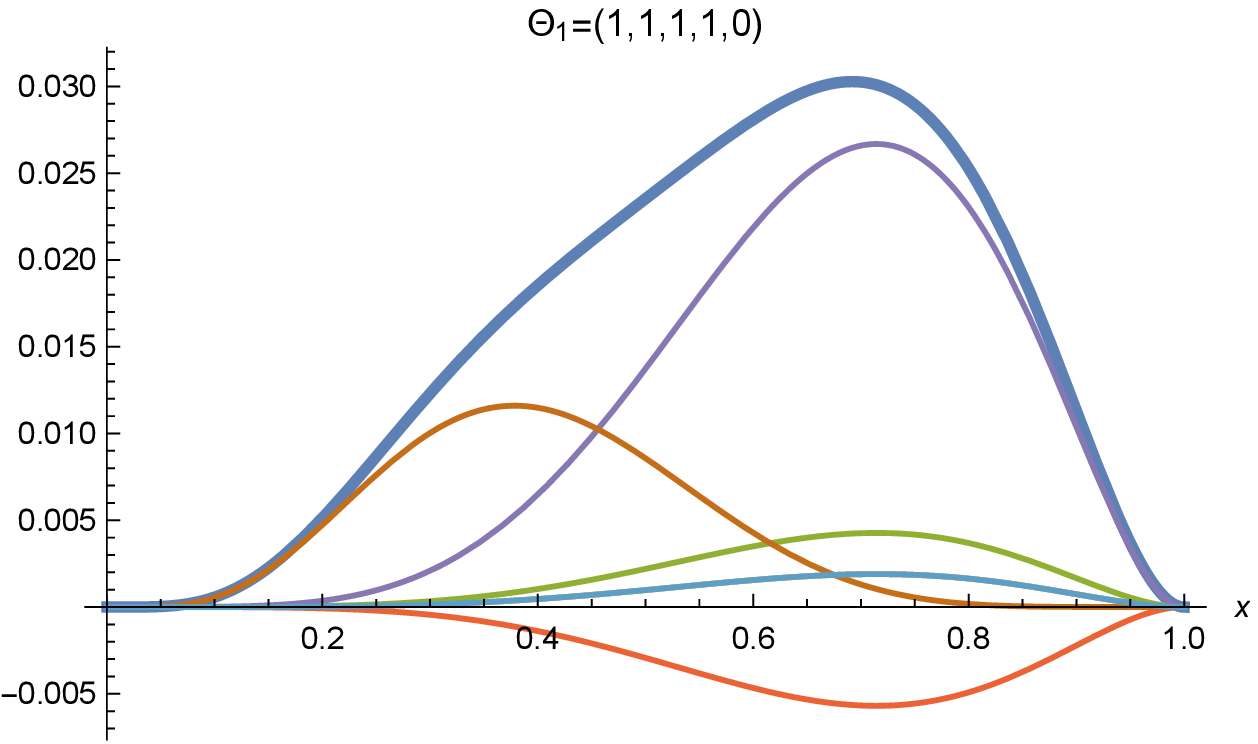}
            \caption[]%
            {{\small $F_1$}}    
        \end{subfigure}
        \hfill
        \begin{subfigure}[b]{0.32\textwidth}   
            \centering 
            \includegraphics[width=\textwidth]{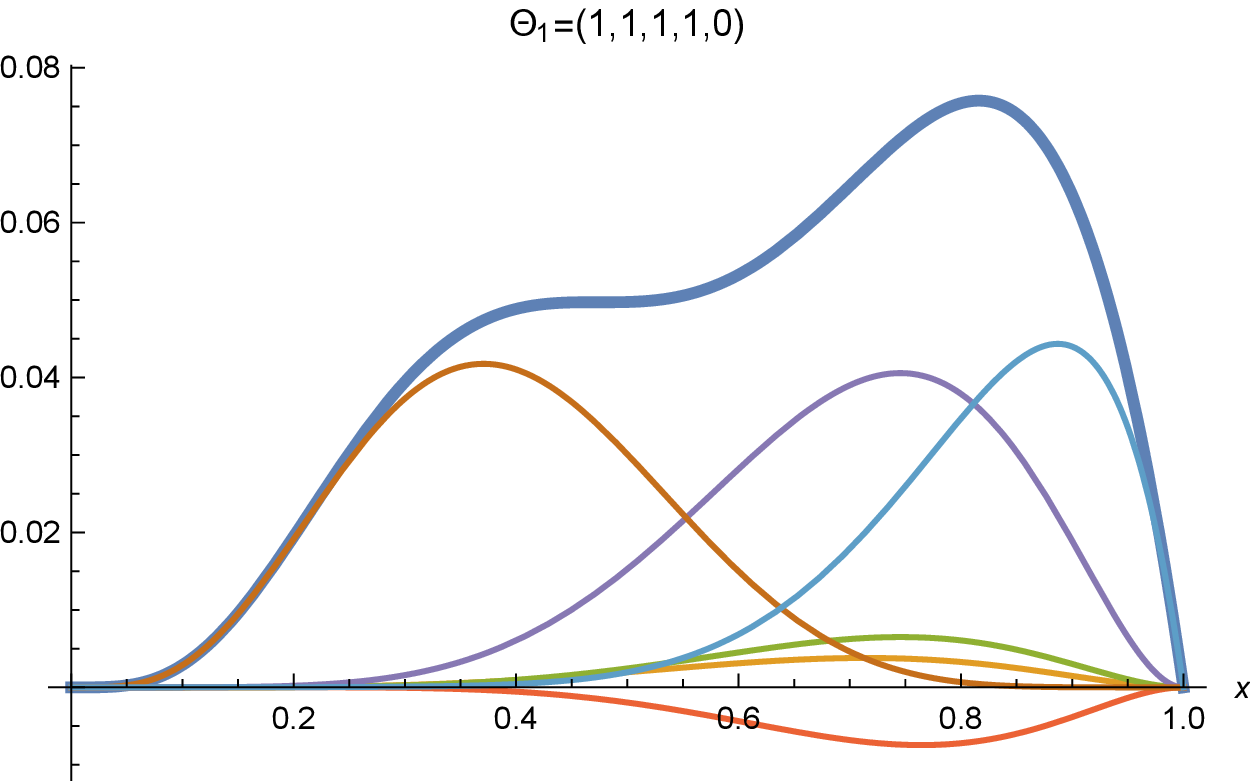}
            \caption[]%
            {{\small $F_2$}}    
        \end{subfigure}
            \hfill
        \begin{subfigure}[b]{0.32\textwidth}  
            \centering 
            \includegraphics[width=\textwidth]{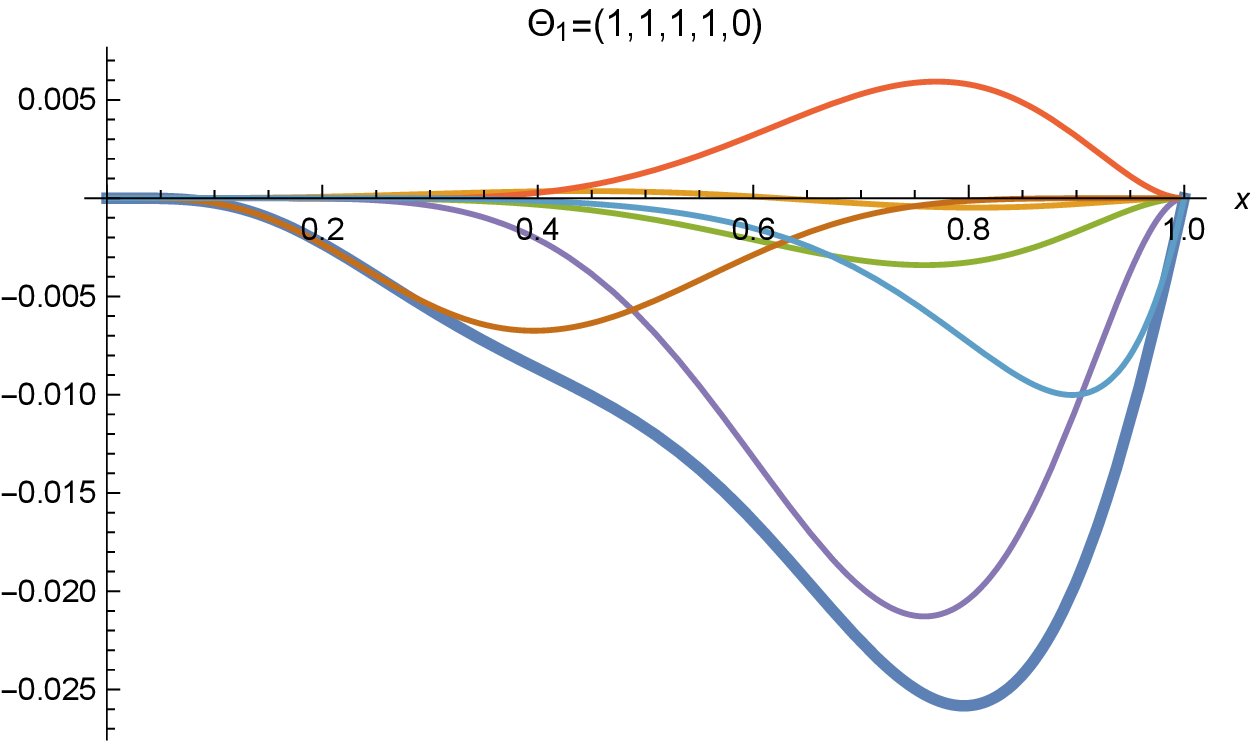}
            \caption[]%
            {{\small $g_2$}}    
        \end{subfigure}
        \vskip\baselineskip
        \caption[ ]
        {\small Full structure functions $F_1$, $F_2$ and $g_2$ (blue line) as functions of the Bjorken parameter $x$, obtained from twist-4 spin-1/2 fermionic operators ${\cal {O}}_{k=1}^{I_1, (6)}$ with $l_5=1\geq l_4\geq l_3\geq l_2\geq l_1=0$ (${\cal{Q}}=1/2$). We also display all the contributions coming from different terms, namely: from the minimal coupling $\beta^2_m F_1^m$ (orange line); the Pauli interaction where the intermediate state $\lambda_{\cal {X}} \equiv \lambda_{k=1}^-$ is the same as the incident state $\beta^2_P F_1^P$ (green line), and where the intermediate state is different  $\beta^2_{Pm} F_1^{P}$ (violet line), both belonging to the {\bf 20$^*$}; the contribution from cross terms $\beta_m \beta_P F_1^c$ (red line); and the contributions from the Pauli interactions with intermediate states in the $\textbf{60}$:  $\beta^2_+ F_1^{P +}$, where $\lambda_{\cal {X}} \equiv \lambda_{k+1}^+$ (brown line); and in the $\textbf{4}$: $\beta^2_- F_1^{P-}$, where $\lambda_{\cal {X}} \equiv \lambda_{k-1}^+$ (light blue line).} 
        \label{fig-tau4-detail}
    \end{figure}

\begin{figure}[H]
        \centering
        \begin{subfigure}[b]{0.32\textwidth}
            \centering
            \includegraphics[width=\textwidth]{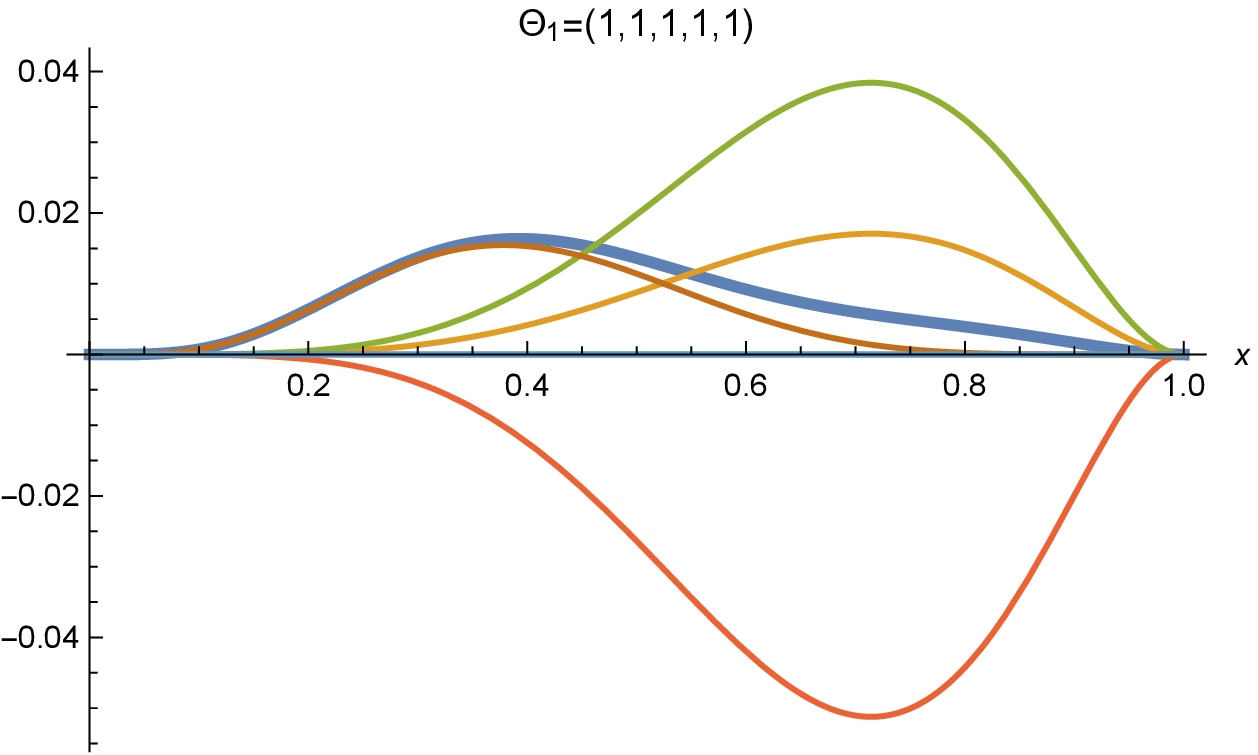}
            \caption[Network2]%
            {{\small $F_1$ }}    
        \end{subfigure}
        \hfill
        \begin{subfigure}[b]{0.32\textwidth} 
            \centering 
            \includegraphics[width=\textwidth]{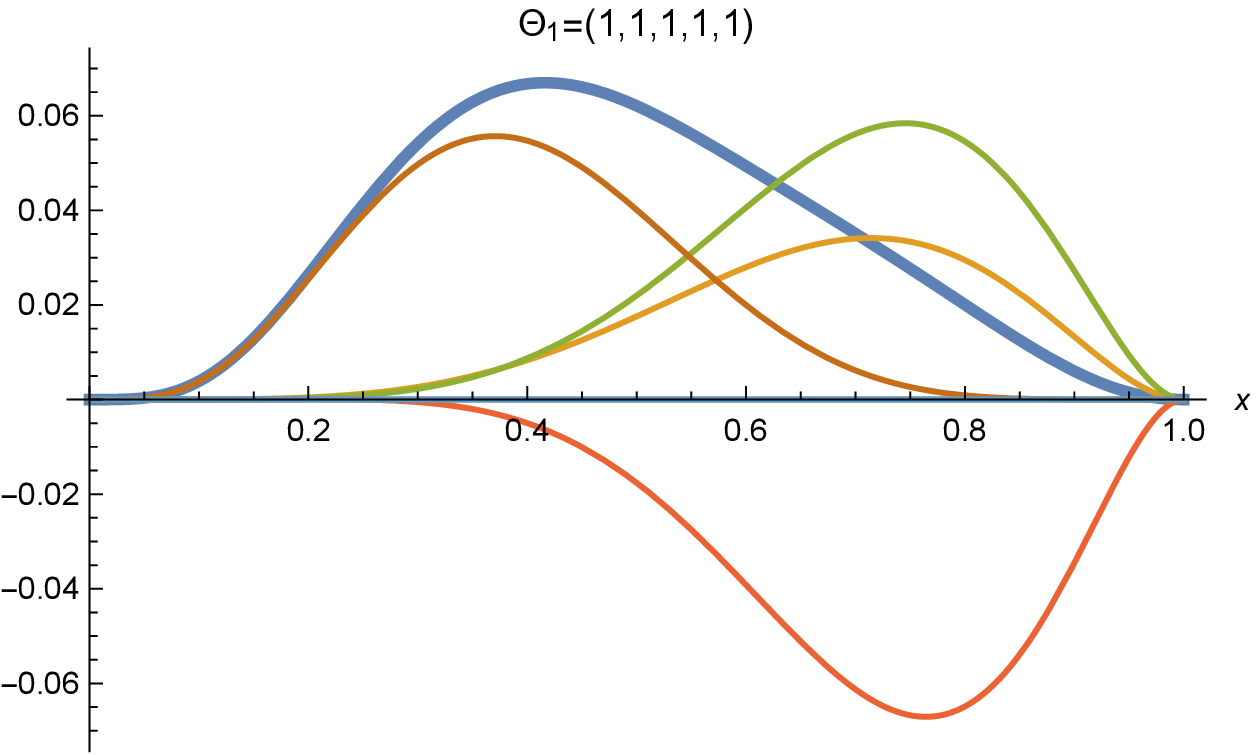}
            \caption[]%
            {{\small $F_2$}}    
        \end{subfigure}
            \hfill
        \begin{subfigure}[b]{0.32\textwidth} 
            \centering 
            \includegraphics[width=\textwidth]{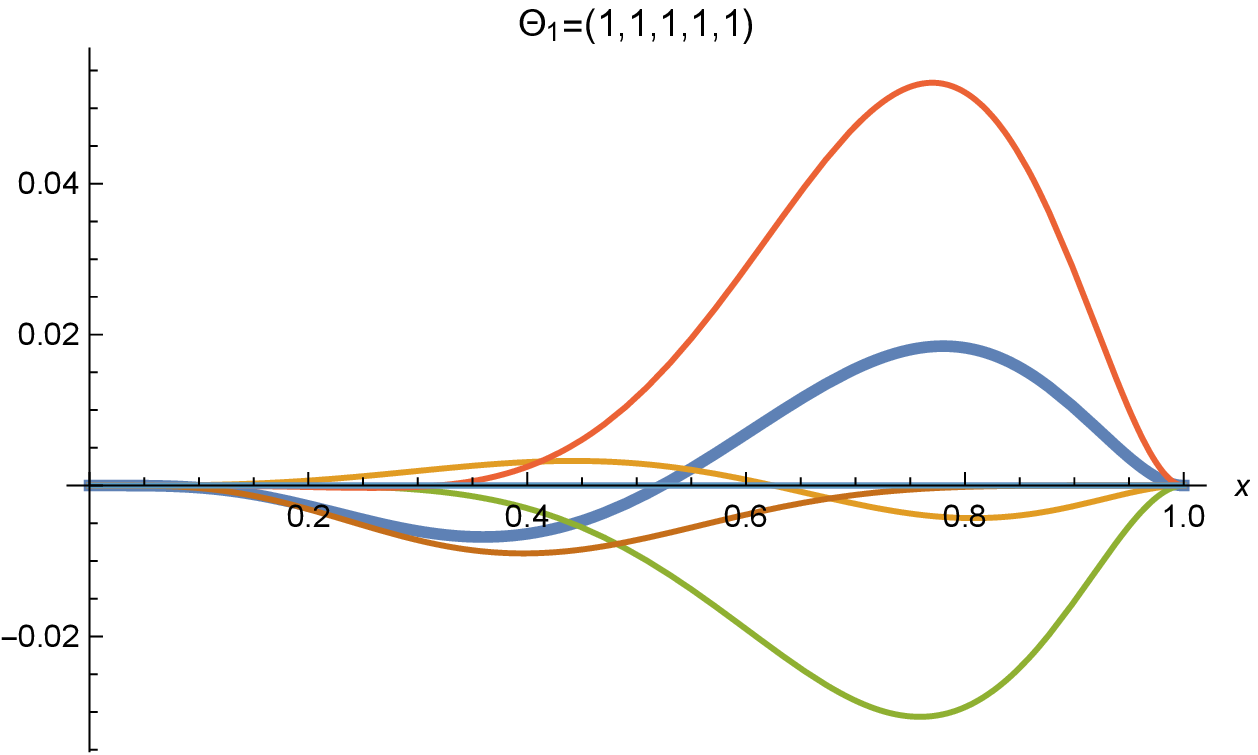}
            \caption[]%
            {{\small $g_2$}}    
        \end{subfigure}
        \caption[ ]
        {\small  Full structure functions $F_1$, $F_2$ and $g_2$ (blue line) as functions of the Bjorken parameter $x$ and the detail of each contributions, obtained from the twist-4 spin-1/2 fermionic operators ${\cal {O}}_{k=1}^{I_1, (6)}$ with $l_5=l_4=l_3=l_2=l_1=1$ (${\cal{Q}}=3/2$). The meaning of the curves is analogous as described in figure 2.} 
        \label{fig-tau4-detail-2}
    \end{figure}

The twists of the incident and the intermediate states, related to the values of $l_5=l=k$, set the dependence on the Bjorken variable as well as on the virtual-photon momentum transfer. Then, the degeneracy given by the rest of numbers $l_4, l_3, l_2$ and $l_1$ can enhance the relative coefficient of a given contribution. In figures \ref{fig-tau4-detail} and \ref{fig-tau4-detail-2} we draw the structure functions $F_1$, $F_2$ and $g_2$ for the five possible incident hadrons  with $(1, l_4, l_3, l_2, l_1)_a$ numbers and $\tau=4$. We have set $|a_0|=1$ which is the only free constant for all the structure functions. In addition we have factorized out $(\Lambda^2/q^2)^{\tau-1}$. In each sub-figure the different contributions of equation (\ref{complete-structure-function}) are displayed with different colors (see figure captions), while blue lines indicate the full structure functions (including all possible contributions). The corresponding curves for $g_3$ are displayed in Appendix \ref{Appendix_C}.

Firstly, we note that for almost all the structure functions, the minimal coupling  contributions (orange line) are very small in comparison with the rest of the contributions. This effect has been observed in \cite{Jorrin:2020cil} for $\tau=3$. However, in the case with maximum charge ${\cal{Q}}=3/2$  ($l_1=1$) they have a similar magnitude in comparison with the other terms as shown in figure 3. Secondly, with respect to the Pauli terms, let us consider the coupling which connects states belonging to the same {\bf 20$^*$} irreducible representation of $SU(4)$  (green and violet lines). They show a bell-shaped form with maximum near $x \sim 0.75$, and also they fall off as $(1-x)^{\tau-2}$ as $x\to1$. These terms are very important for the structure functions with ${\cal{Q}}=1/2$, even taking into account the suppression of the red line given by the cross terms. The behaviour of the violet line is controlled by the set of states $(1, l_4, l_3, l_2, 0)$ (see figure 2), and they do not contribute if the incident hadron has charge ${\cal{Q}}=3/2$ (figure 3).

On the other hand, there are also contributions from the diagrams with intermediate states associated with operators ${\cal{O}}^{I_0, (13)}_{k=0}$ which belong to the $\textbf{4}$ irreducible representation of $SU(4)_R$. The case corresponding to the selection rule $k'=k+1$ (indicated in the figures with brown lines) is interesting since it is relevant for all the structure functions. Particularly, for ${\cal{Q}}=3/2$, this represents the main contribution at relatively low $x$. The corresponding curves have maxima around $x \sim 0.35$ and fall off rapidly for higher values of the Bjorken parameter. Finally, the light-blue curves correspond to the case when $k'=k-1$, and they display their maxima around $x \sim 0.9$. For ${\cal{Q}}=3/2$ charge conservation does not allow for this type of coupling, thus the coefficients obtained from the spinor-spherical-harmonics angular integrals vanish in this case. For ${\cal{Q}}=1/2$ these couplings become negligible for $F_1$, being only relevant for incident hadrons with quantum numbers $(1,1,1,0,0)$ and $(1,1,1,1,0)$ for $F_2$ and $g_2$.

\begin{figure}[H]
        \centering
        \begin{subfigure}[b]{0.45\textwidth}
            \centering 
            \includegraphics[width=\textwidth]{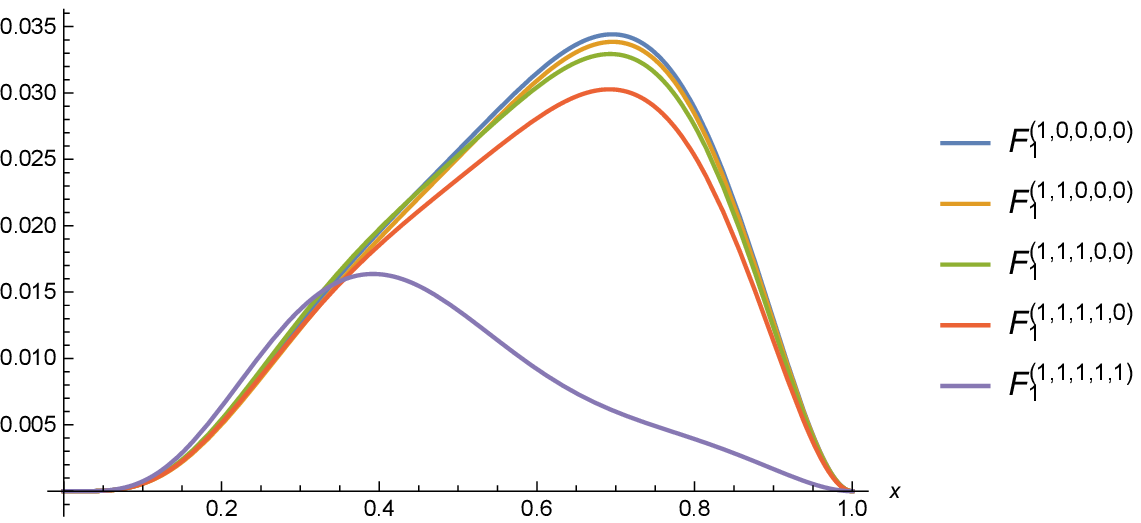}
            \caption[]%
            {{\small $F_1$}}    
        \end{subfigure}
        \hfill
        \begin{subfigure}[b]{0.45\textwidth}
            \centering 
            \includegraphics[width=\textwidth]{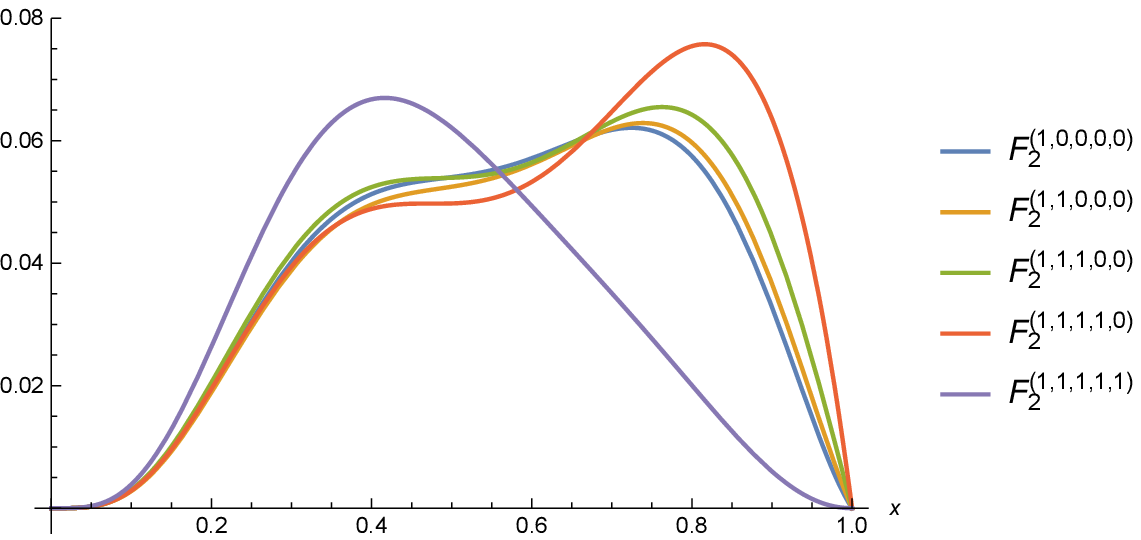}
            \caption[]%
            {{\small $F_2$}}    
        \end{subfigure}
            \hfill
        \begin{subfigure}[b]{0.45\textwidth}
            \centering 
            \includegraphics[width=\textwidth]{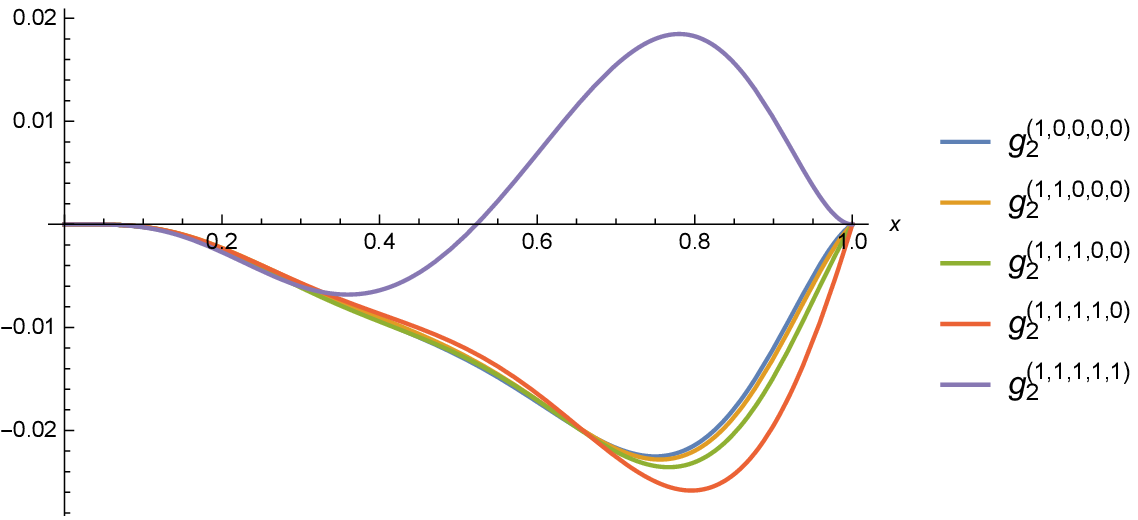}
            \caption[]%
            {{\small $g_2$}}    
        \end{subfigure}
        \caption[ ]
        {\small Full structure functions $F_1$, $F_2$ and $g_2$ as functions of the Bjorken parameter $x$ obtained from the twist-4 spin-1/2 fermionic operators ${\cal {O}}_{k=1}^{I_1,(6)}$ with $(1,l_4,l_3,l_2,l_1)_{a=1}$.} 
        \label{fig-tau4-complete}
    \end{figure}

Figure \ref{fig-tau4-complete} shows the full structure functions for each possible initial state with $\tau=4$. The curves with the same charge (${\cal{Q}}=1/2$) have a similar bell-shaped curves. The region of $0.6 \leq x < 1$ is dominated by contributions associated with the operators ${\cal{O}}^{I_1, (6)}_{k=1}$ as intermediate hadrons.  In contrast, for $0.2\leq x \leq 0.6$ the leading diagram comes from intermediate states associated with the operators ${\cal{O}}^{I_2, (13)}_{k'=2}$ which belong to the {\bf 60} irreducible representation of $SU(4)_R$. The structure functions for states with charge ${\cal{Q}}=3/2$ have a different behaviour, showing a significant suppression of the contribution of intermediate states associated to ${\cal{O}}^{I_1, (6)}_{k=1}$ above $x \sim 0.6$ for $F_1$ and $F_2$. Note that for $g_1$ and $g_2$ the behaviour is different. 


%
\subsection{The case of twist-5 ${\cal {O}}_{k=2}^{I_2, (6)}= C^{I_2, (6)}_{i_1 i_2} \text{Tr}(F_+ \lambda_{{\cal {N}}=4} X_{i_1} X_{i_2})$ operators}
%

\begin{figure}[H]
        \centering
        \begin{subfigure}[b]{0.45\textwidth} 
            \centering 
            \includegraphics[width=\textwidth]{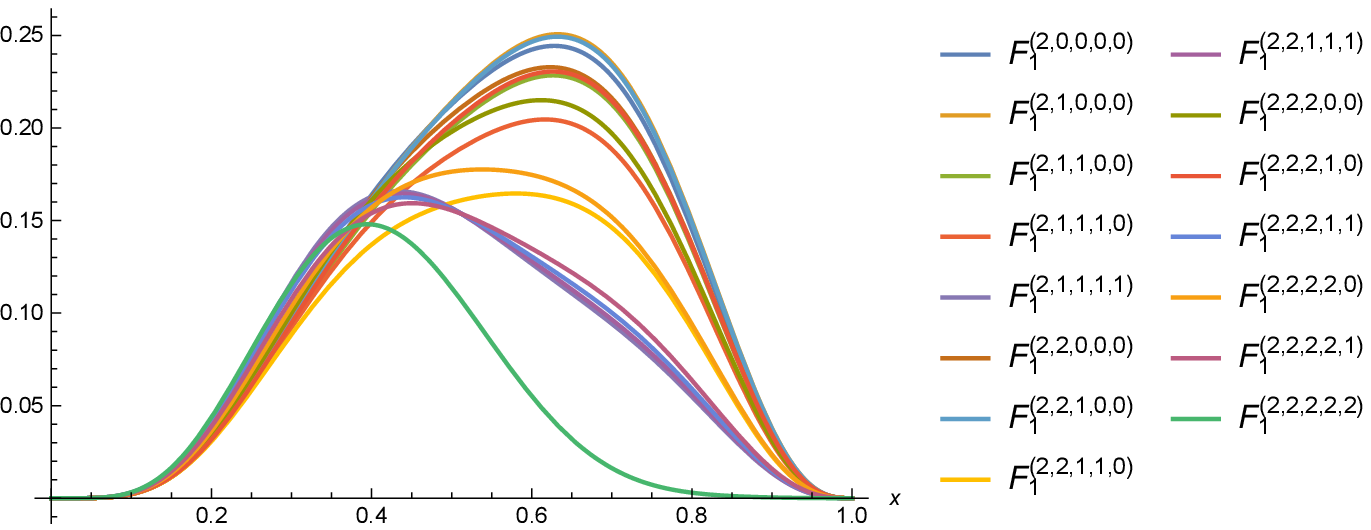}
            \caption[]%
            {{\small $F_1$}}    
        \end{subfigure}
        \hfill
        \begin{subfigure}[b]{0.45\textwidth} 
            \centering 
            \includegraphics[width=\textwidth]{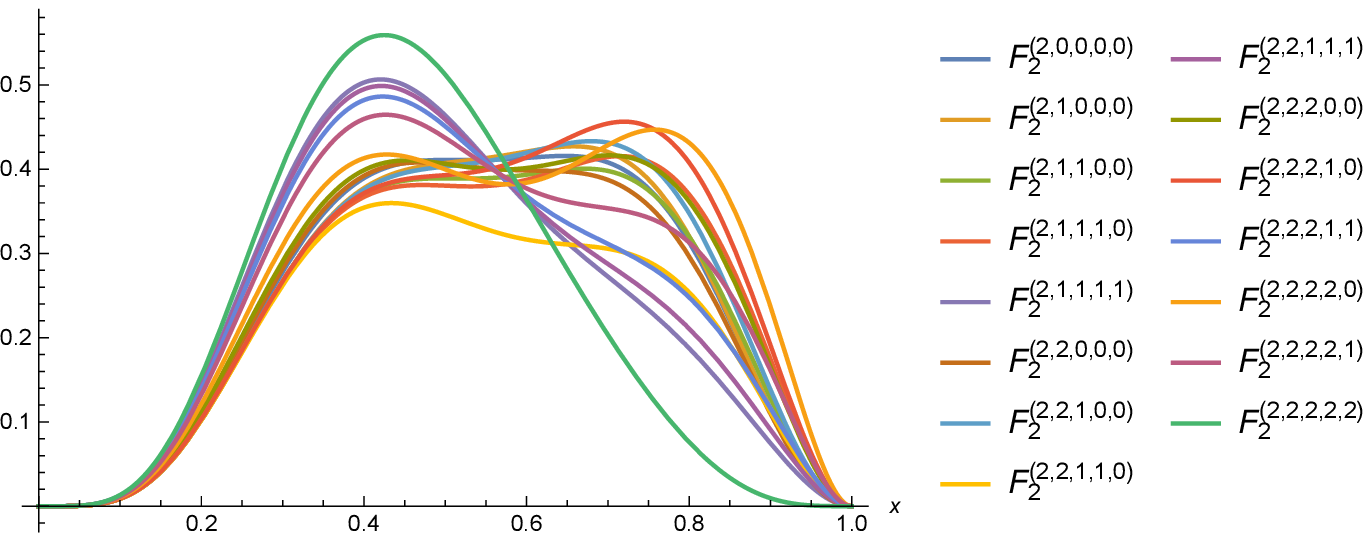}
            \caption[]%
            {{\small $F_2$}}    
        \end{subfigure}
            \hfill
        \begin{subfigure}[b]{0.45\textwidth} 
            \centering 
            \includegraphics[width=\textwidth]{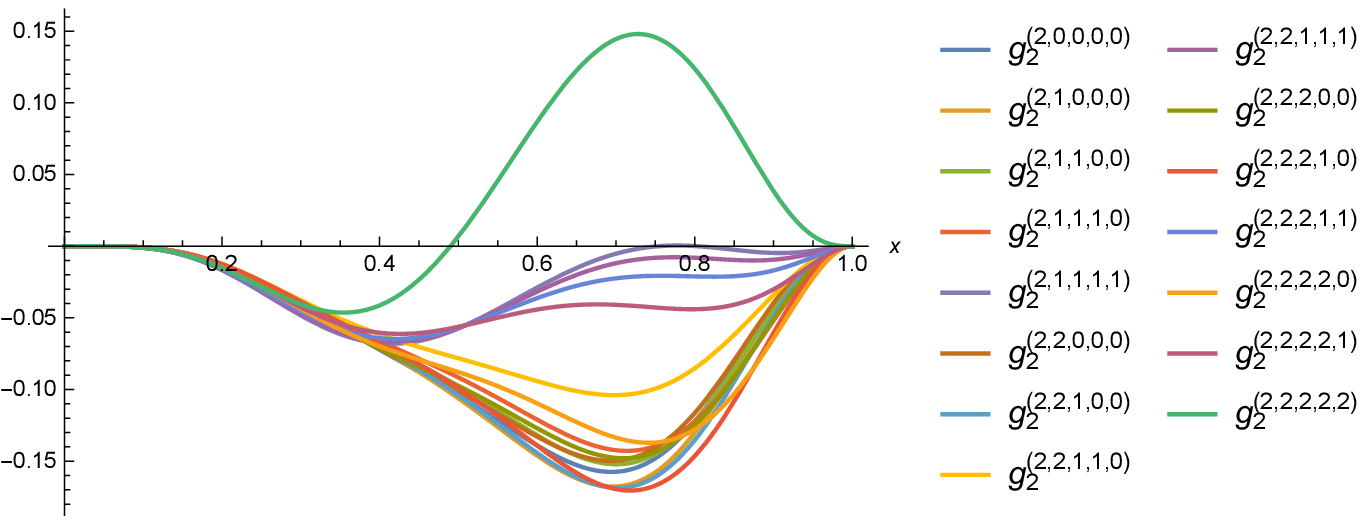}
            \caption[]%
            {{\small $g_2$}}    
        \end{subfigure}
        \caption[]
        {\small Full structure functions $F_1$, $F_2$ and $g_2$ as functions of the Bjorken parameter $x$ obtained from the twist-5 spin-1/2 fermionic operators ${\cal {O}}_{k=2}^{I_2, (6)}$ with $(2,l_4,l_3,l_2,l_1)_{a=1}$.} 
        \label{fig-tau-5}
    \end{figure}

In this subsection we calculate the structure functions for incident hadrons represented by twist $\tau=5$ spin-1/2 fermionic operators ${\cal{O}}^{I_2, (6)}_{k=2}=C^{I_2, (6)}_{i_1 i_2} \text{Tr}(F_+ \lambda_{{\cal {N}}=4} X_{i_1} X_{i_2})$ of the ${\cal {N}}=4$ SYM theory in the planar limit and at strong coupling, for the Bjorken variable within the $\lambda_{SYM}^{-1/2} \ll x < 1$ range. The behaviour of the different interaction terms in the action (\ref{five-dimensional-action}) is very similar to the case $\tau=4$. However, now the number intermediate of hadronic states is much larger. In figure \ref{fig-tau-5} we show the structure functions $F_1$, $F_2$ and $g_3$ for the 15 independent incoming states which can be separated in terms of their charges ${\cal{Q}}=1/2, 3/2, 5/2$ given by $l_1=0, 1, 2$, respectively. There is only one state $(2,2,2,2,2)_a$ with charge ${\cal{Q}}=5/2$. Each of these states have associated four spinors,  $a=1$, 2, 3, 4. They transform in the {\bf 60$^*$} irreducible representation of $SU(4)_R$.

As the value of the charge becomes larger, the Pauli contribution with a maximum in $x \sim 0.7$ decreases. This effect occurs mainly due to two reasons, which are related to the previous case for $\tau=4$. Firstly, the states which have the biggest ${\cal{Q}}$ couple to a smaller number of intermediate states due to charge conservation. They do not even have interactions for ${\cal{Q}}=5/2$. The second reason is related to the charge appearing in the coefficient of the minimal-coupling and the cross terms. The last contribution suppresses the Pauli terms contributions in $F_1$ and $F_2$ and change the curve of the $g_2$ function for the state $(2,2,2,2,2)$.

It is interesting to analyse the characteristic scale corresponding to the structure functions associated with operators with different twists. Figures have been normalized using the same prescription proposed in reference \cite{Jorrin:2020cil} for the minimal $\tau=3$. In this case we have set
\begin{equation}
 |a_0|^2 \left(\frac{\Lambda^2}{q^2}\right)^{\tau-1}  = \pi^2 K^2 4^{\tau-1}c^2_\chi \ c^2_i \left(\frac{\Lambda^2}{q^2}\right)^{\tau-1} =1 \ .   \label{normalisation}
\end{equation}
The idea is to study all the contributions from different couplings corresponding to an operator of a given twist. However, using this normalization the maximum values of the structure functions increase for higher-twist operators due to the factor $\left(\Gamma(\tau)\right)^2$. For instance, the Pauli contribution $F_2^P$ (see Appendix \ref{Appendix-A}) takes the form 
\begin{eqnarray}
  F_2^P(x,q,\tau)\sim  \pi^2 K^2 4^{\tau-1}c^2_\chi c^2_i \left(\frac{\Lambda^2}{q^2}\right)^{\tau-1} \left(\Gamma(\tau)\right)^2 x^{\tau+1}(1-x)^{\tau-2} [1+\tau(\tau-2) \ x] \ ,
\end{eqnarray}
where the factor $\left(\Gamma(\tau)\right)^2$ scales the contributions, and it increases for higher twists, considering  equation (\ref{normalisation}).

In order to clarify this behaviour, let us recall that the normalization of the structure functions depends on $\tau$ and the calculation has been carried out in the DIS limit ($q\sim M_\chi \gg M_i\sim \Lambda_{QCD}$). The standard normalization used in \cite{Polchinski:2002jw,Mamo:2019mka} implies that the constant $C$ in equation (\ref{spinor-pos}) is written in terms of the twist $\tau$ and the hadronic mass $M_{\tau,i}$ as follows 
\begin{equation}
C=\frac{\sqrt{2} }{z_0 \ J_{\tau-1}(Mz_0)}. 
\end{equation}
Thus, the normalization constant of the incident hadrons is $c_i=\frac{\sqrt{2} (M_i z_0)^{\tau-2} }{J_{\tau-1}(M_i z_0) 2^{\tau-2} \Gamma(\tau-1)}$, while for the intermediate state, by using the asymptotic expansion of the Bessel function around $M_\chi z_0\gg1$, we obtain $c_\chi\sim \pi$. Finally, we can calculate the structure function $F_2^P$ evaluated in the maximum $x_{max}=\frac{1+\tau}{2\tau-1}$ with the full $\tau$-dependence of the normalization constants, obtaining
\begin{eqnarray}
F_2^P(x_{max},q,\tau) &\sim &  \frac{K^2 \tau^2}{(M_i z_0)^2 \left(J_{\tau-1}(M_i z_0)\right)^2} \left(\frac{M^2_i}{q^2}\right)^{\tau-1} x_{max}^{\tau+1}(1-x_{max})^{\tau-2} (1+x_{max}\tau(\tau-2))  , \nonumber \\
&&
\end{eqnarray}
where $M_i$ is the mass of the incident hadron and it depends on the $j$-th zero of the Bessel function\footnote{For each $\tau$, there is a mass tower of hadrons $M_{\tau,j}$ which is given by the $j$-th zero  of the Bessel functions $J_{\tau-2}(R_{\tau,j})=0$ (hard-wall model), where $R_{\tau,j} = M_{\tau,j} z_0$.}. The maximum value of the contribution scales with $\left(M_i^2/q^2\right)^{\tau-1}$. Thus, it becomes evident how the fall off is accentuated for higher-twist operators. For example, phenomenologically we can consider the kinematic variables taking the values $q^2 \sim 15$ GeV$^2$, $M_i \sim 1$ GeV and $\Lambda_{QCD}\sim 0.2$ GeV. Therefore, if we consider the lower hadronic mass for each $\tau$, the maximum value for $\tau=3$ is approximately 8.6 times greater than the case $\tau=4$, and 47 times greater than the case $\tau=5$. These results may be important to compare with the QCD phenomenology
\cite{QCD-new}.

%
\section{Comments on the results and conclusions}
%


We have done an exhaustive study of different contributions to the structure functions of electromagnetic DIS off polarised spin-1/2 hadrons using the AdS/CFT duality. Particularly, we have focused on local single-trace higher-twist spin-1/2 fermionic operators of the strongly coupled $SU(N)$ ${\cal {N}}=4$ SYM theory in the planar limit. Specifically, we have worked out in full detail the cases with $\tau=4$ and 5. It is worth mentioning that we have carried out all our calculations from first principles, {\it i.e.} considering the background AdS$_5 \times S^5$ from type IIB supergravity. In the effective five-dimensional action (\ref{five-dimensional-action}) there are several contributions. The first one comes from the minimal-coupling term. In addition, there are very important contributions from the second and third terms which are Pauli interactions. For all structure functions the contributions obtained from the minimal-coupling term turn out to be smaller in comparison with those emerging from the different Pauli terms. This effect is manifested on both the symmetric as well as the anti-symmetric structure functions, as shown in all the figures displayed in this work. In comparison with the case of ${\cal {O}}_0^{I_0, (6)}$ operators in the ${\bf 4^*}$ irreducible representation of $SU(4)_R$, which are only four different operators, an consequently there are only four different dual Kaluza-Klein dilatino modes, in this work for $\tau=4$, the operators are in the ${\bf 20^*}$ irreducible representation of $SU(4)_R$ and the dual corresponding fermionic type IIB supergravity modes are also twenty.  For $\tau=5$ these figures become 60 SYM operators and their corresponding 60 Kaluza-Klein states, respectively. On the other hand, calculations become much more complicated for twists 4 and 5 in comparison with 3. In addition, there are new relevant terms from the selection rule $\lambda_k \rightarrow \lambda_{k-1}$ with $k>0$. All this has been discussed in detail in the previous sections.

Another interesting point to mention concerns the OPE of two electromagnetic currents inside the hadron. Its matrix elements define the tensor (\ref{Ttensor}), that can be expanded in terms of the generic structure functions $\tilde{F}_i(x, q^2/\Lambda^2)$. The optical theorem leads to equations (\ref{TW}), thus the relation with the usual structure functions defining the hadronic tensor $W_{\mu\nu}$ is $F_i(x, q^2/\Lambda^2)\equiv 2 \pi \text{Im} \tilde{F}_i(x, q^2/\Lambda^2)$. Then, we can obtain the moments of the structure functions $M_n^{i}(q^2)$, which can be schematically written as the sum of three kind of contributions \cite{Polchinski:2002jw}. The leading contribution at weak coupling comes from twist-2 operators. In reference \cite{Kotikov:2002ab} the DGLAP and BFKL evolution equations in the ${\cal {N}}=4$ SYM theory at the next-to-leading approximation have been derived. Also, the evaluation of Wilson coefficients for DIS has been done in the next-to-leading approximation in \cite{Bianchi:2013sta}. There are other contributions to the moments of the structure functions $M_n^{i}(q^2)$ from the non-perturbative domain, which are the ones we have considered in the present work. These contributions come from double-trace operators constructed from protected single-trace operators. In the present case these are the protected single-trace twist-$\tau$ (for $\tau \geq 3$) spin-1/2 operators of ${\cal {N}}=4$ SYM theory at strong coupling of the form ${\cal {O}}_k^{I_k, (6)}$ which belong to the {\bf 4$^*$}, {\bf 20$^*$}, {\bf 60$^*$}, $\cdots$ irreducible representations of $SU(4)_R$, labelled as $[1, k, 0]$. As we have seen in previous sections there are also contributions from operators of the form ${\cal {O}}_k^{I_k, (13)}$ in the {\bf 4}, {\bf 20}, {\bf 60}, $\cdots$ irreducible representations of $SU(4)_R$, labelled as $[0, k, 1]$. These contributions to the OPE are the leading ones in the large-$N$ limit and correspond to final single-particle states in DIS or the exchange of single-particle intermediate states in the forward Compton scattering. In terms of the type IIB supergravity dual description these processes only involve a single intermediate dilatino mode exchange. This is what we have developed in this work.

There is a third kind of contributions which are relevant for finite $N$, and they correspond to the exchange of two or more particle intermediate states in the forward Compton scattering. 
They correspond to multi-trace operators in the SYM theory.
In the present case we have not considered $1/N$ corrections since here we focus on the large-$N$ limit of the $SU(N)$ ${\cal {N}}=4$ SYM theory. We have obtained the structure functions for exchange of two-particle intermediate states for glueballs in \cite{Jorrin:2016rbx} for ${\cal {N}}=4$ SYM theory in terms of type IIB supergravity on AdS$_5 \times S^5$, scalar mesons in \cite{Kovensky:2016ryy} and vector mesons in \cite{Kovensky:2018gif} both in the context of the D3D7-brane model \cite{Kruczenski:2003be}.

The techniques presented in this work can be used to study other single-trace operators of the $SU(N)$ ${\cal {N}}=4$ SYM theory. Also, it would be very interesting to extend it to study cases with less supersymmetries such as ${\cal {N}}=1$ SYM theory developed by Klebanov and Witten \cite{Klebanov:1998hh}, considering the spectrum of type IIB supergravity on AdS$_5 \times T^{1,1}$ \cite{Ceresole:1999zs,Ceresole:1999rq}. In this case the angular integrals on $T^{1,1}$ would set different selection rules which likely induce new interesting effects. In similar lines it would be interesting to investigate the case of
type I' string theory associated with five-dimensional supersymmetric fixed points with $E_{N_{f+1}}$ global symmetry.
This includes many different theories with $E_{N_{f +1}} = E_8, E_7$, $E_6$, $E_5 = Spin(10)$, $E_4 = SU(5)$, $E_3 = SU(3) \times SU(2)$, $E_2 = SU(2) \times U(1)$, $E_1 = SU(2)$ global symmetry groups \cite{Seiberg:1996bd}. The gravity duals of these theories were constructed in \cite{Brandhuber:1999np}, and are related to the near horizon limit of the D4D8-brane system in massive type IIA supergravity \cite{Romans:1985tz} compactified on the AdS$_6 \otimes S^4$ fibration \cite{DAuria:2000afl}. Interesting related gauge/supergravity duals have been obtained in \cite{Nunez:2001pt}.
Also, using the ideas discussed above it would be interesting to investigate eleven-dimensional supergravity on AdS$_4 \times S^7$ and AdS$_7 \times S^4$ \cite{Ferrara:1999bv,Ferrara:2000dv,Sezgin:2020avr} as well as deformations leading to SYM theories preserving less supersymmetries \cite{Gursoy:2002tx}.

~

~

%
\centerline{\large{\bf Acknowledgments}}
%

~

We thank Gustavo Michalski for collaboration in early stages of the project and for a critical reading of the manuscript.
This work has been supported in part by the National Scientific Research Council of Argentina (CONICET), the National Agency for the Promotion of Science and Technology of Argentina (ANPCyT-FONCyT) Grants PICT-2015-1525 and PICT-2017-1647, the UNLP Grant PID-X791, and the CONICET Grants PIP-UE B\'usqueda de nueva f\'{\i}sica and PICT-E 2018-0300 (BCIE).

\newpage

\appendix

\section{Minimal coupling and Pauli terms  contributions to the structure functions} \label{Appendix-A}

In this Appendix we introduce the structure functions associated with each interaction term which are used to draw the figures for higher-twist spin-1/2 operators. For the minimal coupling the structure functions of a incident polarized spin-1/2 hadron with twist $\tau$ are given by
\begin{eqnarray}
F^m_1&=&\frac{F_2^m}{2}=\frac{F_3^m}{2}=g_1^m=\frac{g_3^m}{2}=\frac{g_4^m}{2}=\frac{g_5^m}{2} 
%
%
=\frac{|a_0|^2}{8} \ \Gamma^2(\tau)  \left(\frac{\Lambda^2}{q^2}\right)^{\tau-1} x^{\tau+1}(1-x)^{\tau-2}  \, , \nonumber  \\
&& \\
g_2^m&=& \left( \frac{1}{2}\frac{\tau+1}{\tau-1}-\frac{x \tau}{\tau-1} \right) \frac{|a_0|^2}{8} \ \Gamma^2(\tau) \left(\frac{\Lambda^2}{q^2}\right)^{\tau-1} x^{\tau}(1-x)^{\tau-2} \, ,
\end{eqnarray}
where $\Gamma(x)$ is the Gamma function and $|a_0|=2 \pi c'_i c'_{\cal {X}}2^\tau K$ is a constant.

For the Pauli interactions between states in the same representation the structure functions are:
\begin{eqnarray}
F^P_1&=&\frac{F_3^P}{2}=g_1^P=\frac{g_5^P}{2}=\frac{1}{2} |a_0|^2 \ \Gamma^2(\tau) \left(\frac{\Lambda^2}{q^2}\right)^{\tau-1} x^{\tau+1}(1-x)^{\tau-2} (1-\tau)^2 \, , \\
F^P_2&=&g_4^p= |a_0|^2 \ \Gamma^2(\tau) \left(\frac{\Lambda^2}{q^2}\right)^{\tau-1}  x^{\tau+1}(1-x)^{\tau-2} (1+x\tau(\tau-2)) \, , \nonumber \\
\\
g_2^P&=& - \frac{1}{4} |a_0|^2 \ \Gamma^2(\tau) \left(\frac{\Lambda^2}{q^2}\right)^{\tau-1}  (1-x)^{\tau-2} x^{\tau+1}(\tau(1-\tau+x(3+2(\tau-2)\tau))-1)/(\tau-1), \nonumber \\
&& \\
g_3^P&=& - |a_0|^2 \ \Gamma^2(\tau) \left(\frac{\Lambda^2}{q^2}\right)^{\tau-1} (1-x)^{\tau-2} x^{\tau+1}(1+\tau (1-3x)+(2x-1)\tau^2)/(\tau-1) \, .
\end{eqnarray}
Finally, the structure functions from the cross-terms contribution having both the minimal coupling and the Pauli interactions are
\begin{eqnarray}
F^{c}_1&=&\frac{F_3^{c}}{2}=g_1^c=\frac{g_5^c}{2}=\frac{1}{2} |a_0|^2 \left(\frac{\Lambda^2}{q^2}\right)^{\tau-1}  (1 - x)^{\tau-2} x^{\tau+1} (\tau-1) \Gamma(\tau)^2  \, , \\
F^{c}_2&=&g_4^c= |a_0|^2 \left(\frac{\Lambda^2}{q^2}\right)^{\tau-1} (1 - x)^{\tau-2} x^{\tau+1} (-1 + x\tau) \Gamma(\tau)^2 \, ,
\\
g_2^{c}&=&- \frac{1}{4} |a_0|^2 \left(\frac{\Lambda^2}{q^2}\right)^{\tau-1}\frac{\Gamma(\tau)^2}{\tau-1} (1-x)^{\tau-2} x^{\tau+2} (2-\tau^2+x\tau(4\tau-5)) \, ,
\\
g_3^{c}&=& \frac{1}{2} \frac{\Gamma(\tau)^2}{\tau-1} |a_0|^2 \left(\frac{\Lambda^2}{q^2}\right)^{\tau-1} (1-x)^{\tau-2} x^{\tau+1} \left(2-(4+x)\tau +(-1+4x)\tau^4 \right) \, .
\end{eqnarray}

\section{Tables of angular integrals}
\label{Appendix_B}

The coefficients corresponding to the terms in the action (\ref{five-dimensional-action}) are calculated from the angular integrals of the spinor spherical harmonics and the Killing vectors $v^{\alpha}$. The results of the following integrals are shown in table \ref{table_1},
\begin{eqnarray}
    \int d \Omega_5 (\Theta^-_{(1,l'_4,l'_3,l'_2,l'_1)_a})^*  v^i \tau_i \Theta^-_{(1,l_4,l_3,l_2,l_1)_{a=1}} \, ,
\end{eqnarray}
where the incoming spinor spherical harmonics is $\Theta^-_{(1,l_4,l_3,l_2,l_1)_{a=1}}$ while the outgoing one is given by $\Theta^-_{(1,l'_4,l'_3,l'_2,l'_1)_a}$. In this case both of them belong to the same Kaluza-Klein mass tower.

Then, in tables \ref{table_2} and \ref{table_3} are listed the results of the following integrals between states belonging to different Kaluza-Klein mass towers,
\begin{eqnarray}
    \int d \Omega_5 (\Theta^+_{(1\pm1,l'_4,l'_3,l'_2,l'_1)_a})^*  v^i \tau_i \Theta^-_{(1,l_4,l_3,l_2,l_1)_{a=1}} \, .
\end{eqnarray}
In this case the outgoing spinors have superscript (+) and $l_5$ can only take the values $1\pm1$.

\begin{table}[H]
\def\arraystretch{1.5}
\begin{center}
\begin{tabular}{|c|c|c|c|c|c|} 
%
 \hline
 & $(1,0,0,0,0)^-_{1}$ & $(1,1,0,0,0)^-_{1}$  & $(1,1,1,0,0)^-_{1}$ & $(1,1,1,1,0)^-_{1}$ & $(1,1,1,1,1)^-_{1}$  \\
\hline
$(1,0,0,0,0)^-_{1}$  & $-\frac{7}{30}$  & $-\frac{1}{20 \sqrt{6}}$   &0  &0  &0    \\
\hline
$(1,0,0,0,0)^-_{3}$ & 0 &0   & $\frac{1}{12\sqrt{10}}$  & $\frac{1}{12\sqrt{5}}$   &0  \\
\hline
$(1,1,0,0,0)^-_{1}$ & -$\frac{1}{20 \sqrt{6}}$& $-(9/40)$ &0 &0  &0   \\
\hline
$(1,1,0,0,0)^-_{3}$ & 0 & 0 &$\frac{1}{8\sqrt{15}}$  & $\frac{1}{4\sqrt{30}}$&0    \\
\hline
$(1,1,1,0,0)^-_{1}$ & 0& 0 &$-\frac{5}{24}$  & $\frac{1}{12\sqrt{2}}$  &0   \\
\hline
$(1,1,1,0,0)^-_{3}$ & -$\frac{1}{12\sqrt{10}}$ &$-\frac{1}{8\sqrt{15}}$  &0 &0  &0   \\
\hline
$(1,1,1,1,0)^-_{1}$ & 0  & 0&$\frac{1}{12\sqrt{2}}$  & $-\frac{1}{6}$ & 0 \\
\hline
$(1,1,1,1,0)^-_{3}$ & $\frac{1}{12\sqrt{5}}$ &$-\frac{1}{4\sqrt{30}}$  & 0  &0  &0    \\
\hline
$(1,1,1,1,1)^-_{1}$ & 0 &  0&0 &0  & $-\frac{1}{2}$   \\
\hline
$(1,1,1,1,1)^-_{3}$ & 0 & 0 &0  &0  & 0  \\
\hline
\end{tabular}
\caption{\small  Results of the angular integrals between states belonging to the same Kaluza-Klein mass tower. The spinor spherical harmonics corresponding to the incoming states are indicated in the first row. The outgoing states are listed in the first column.}
\label{table_1}
\end{center}
\end{table}

\begin{table}[H]
\def\arraystretch{1.5}
\begin{center}
\begin{tabular}{|c|c|c|c|c|c|}
\hline
%
 & $(1,0,0,0,0)^-_{1}$ & $(1,1,0,0,0)^-_{1}$  & $(1,1,1,0,0)^-_{1}$ & $(1,1,1,1,0)^-_{1}$ & $(1,1,1,1,1)^-_{1}$  \\
\hline
$(0,0,0,0,0)^+_{1}$  & -$\frac{1}{3\sqrt{5}}$  & $\frac{1}{\sqrt{30}}$  & 0 & 0 & 0  \\
\hline
$(0,0,0,0,0)^+_{3}$ & 0 & 0  & $-\frac{1}{3\sqrt{2}}$  & $-\frac{1}{3}$  & 0\\
\hline
$(2,0,0,0,0)^+_{1}$ &$\frac{\sqrt{3}}{10}$  &$-\frac{1}{20\sqrt{2}}$    & 0&0  & 0   \\
\hline
$(2,0,0,0,0)^+_{3}$ & 0 &0   & $\frac{1}{4\sqrt{30}}$  &$\frac{1}{4\sqrt{15}}$   &  0  \\
\hline
$(2,1,0,0,0)^+_{1}$ & $\frac{\sqrt{21}}{\sqrt{2} \ 20}$ & $\frac{9}{40\sqrt{7}}$   &0 &0  & 0   \\
\hline
$(2,1,0,0,0)^+_{3}$ &0  &0   & $-\frac{1}{8\sqrt{105}}$  &$-\frac{1}{4\sqrt{210}}$   &0    \\
\hline
$(2,1,1,0,0)^+_{1}$ & 0 & 0  & $\frac{5}{24\sqrt{7}}$  &  $-\frac{1}{12\sqrt{14}}$  &   0 \\
\hline
$(2,1,1,0,0)^+_{3}$ & $\frac{\sqrt{7}}{4\sqrt{10}}$ & $-\frac{1}{8\sqrt{105}}$   &0 &0  &0    \\
\hline
$(2,1,1,1,0)^+_{1}$ & 0& 0  &$-\frac{1}{12\sqrt{14}}$  &$\frac{1}{6\sqrt{7}}$  &  0  \\
\hline
$(2,1,1,1,0)^+_{3}$ & $-\frac{\sqrt{7}}{4\sqrt{5}}$ & $\frac{1}{4\sqrt{210}}$   &0 &0  &0    \\
\hline
$(2,1,1,1,1)^+_{1}$ & 0 &0   &0 &0  & $\frac{1}{2\sqrt{7}}$   \\
\hline
$(2,1,1,1,1)^+_{3}$ &  0&  0 &0 & 0 & 0   \\
\hline
$(2,2,0,0,0)^+_{1}$ &  0&$-\frac{\sqrt{5}}{4\sqrt{7}}$    &0 &0  &  0  \\
\hline
$(2,2,0,0,0)^+_{3}$ &  0& 0  & $\frac{1}{4\sqrt{21}}$  &$\frac{1}{2\sqrt{42}}$   &  0  \\
\hline
$(2,2,1,0,0)^+_{1}$ &  0& 0  & $\frac{5}{12\sqrt{21}}$  & $-\frac{1}{6\sqrt{42}}$  &  0  \\
\hline
$(2,2,1,0,0)^+_{3}$ &  0& $\frac{\sqrt{5}}{4\sqrt{7}}$   &0 &0  &0    \\
\hline
$(2,2,1,1,0)^+_{1}$ &  0& 0  & $\frac{1}{6\sqrt{42}}$  &$-\frac{1}{3\sqrt{21}}$   & 0   \\
\hline
$(2,2,1,1,0)^+_{3}$ &  0& $\frac{\sqrt{5}}{2\sqrt{14}}$   &0 &0  &0    \\
\hline
$(2,2,1,1,1)^+_{1}$ &  0& 0  &0 &0  &$\frac{1}{\sqrt{21}}$    \\
\hline
$(2,2,1,1,1)^+_{3}$ &  0&  0 & 0&0  & 0   \\
\hline
$(2,2,2,0,0)^+_{1}$ &  0& 0&0 &0  &  0  \\
\hline
$(2,2,2,0,0)^+_{3}$ &  0& 0  & $\frac{\sqrt{2}}{3\sqrt{3}}$  &$\frac{1}{6\sqrt{3}}$   &  0 \\
\hline
\end{tabular}
\caption{\small 
Results of the angular integrals between states belonging to different Kaluza-Klein mass towers. In the first row there are the incoming spinor spherical harmonics while the outgoing states are listed in the first column.
}\label{table_2}
\end{center}
\end{table}

\begin{table}[H]
\def\arraystretch{1.5}
\begin{center}
\begin{tabular}{|c|c|c|c|c|c|}
\hline
 & $(1,0,0,0,0)^-_{1}$ & $(1,1,0,0,0)^-_{1}$  & $(1,1,1,0,0)^-_{1}$ & $(1,1,1,1,0)^-_{1}$ & $(1,1,1,1,1)^-_{1}$  \\
\hline
$(2,2,2,1,0)^+_{1}$ &  0& 0  &0 &0  & 0  \\
\hline
$(2,2,2,1,0)^+_{3}$ &  0& 0  & $-\frac{\sqrt{5}}{3\sqrt{6}}$  &$\frac{1}{3\sqrt{15}}$   &    0\\
\hline
$(2,2,2,1,1)^+_{1}$ &  0& 0  &0 &0  & 0   \\
\hline
$(2,2,2,1,1)^+_{3}$ &  0& 0 &0 & 0 & $\frac{1}{\sqrt{15}}$   \\
\hline
$(2,2,2,2,0)^+_{1}$ &  0& 0  &0 & 0 &0   \\
\hline
$(2,2,2,2,0)^+_{3}$ &  0& 0 &0 & $\frac{\sqrt{3}}{2\sqrt{5}}$  &  0 \\
\hline
$(2,2,2,2,1)^+_{1}$ &  0& 0&0 & 0 & 0 \\
\hline
$(2,2,2,2,1)^+_{3}$ &  0& 0&0 & 0 & $-\frac{1}{\sqrt{10}}$  \\
\hline
$(2,2,2,2,2)^+_{1}$ &  0&0 &0 & 0 &0   \\
\hline
$(2,2,2,2,2)^+_{3}$ &  0&0 &0 & 0 & 0  \\
\hline
\end{tabular}
\caption{\small 
Results of the angular integrals between states belonging to different Kaluza-Klein mass towers. In the first row there are the incoming spinor spherical harmonics while the outgoing states are listed in the first column.}\label{table_3}
\end{center}
\end{table}

\section{Results of the structure function $g_3$}
\label{Appendix_C}

In this appendix we show the structure function $g_3$ for $\tau=3$ and $\tau=4$. 

\begin{figure}[H]
        \centering               \begin{subfigure}[b]{0.45\textwidth} 
            \centering 
            \includegraphics[width=\textwidth]{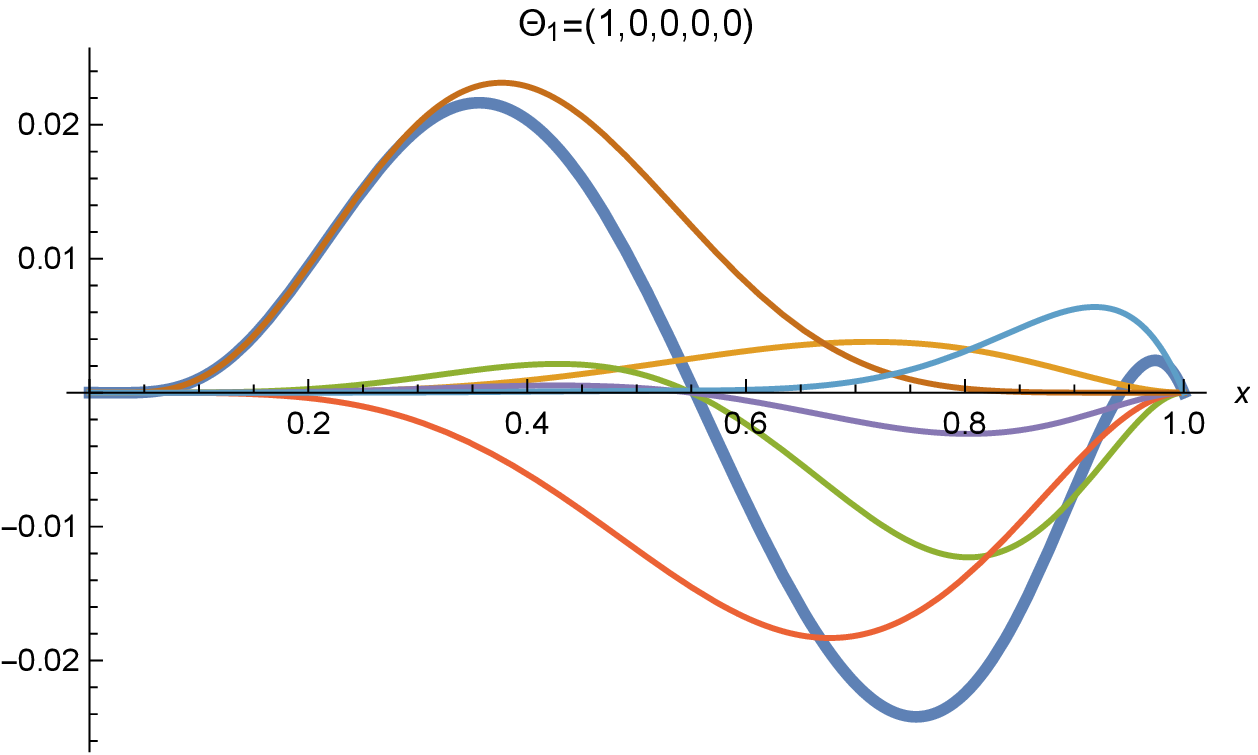}
            \caption[]%
            {{\small $g_3$}}    
        \end{subfigure}
        \hfill
        \begin{subfigure}[b]{0.45\textwidth} 
            \centering 
            \includegraphics[width=\textwidth]{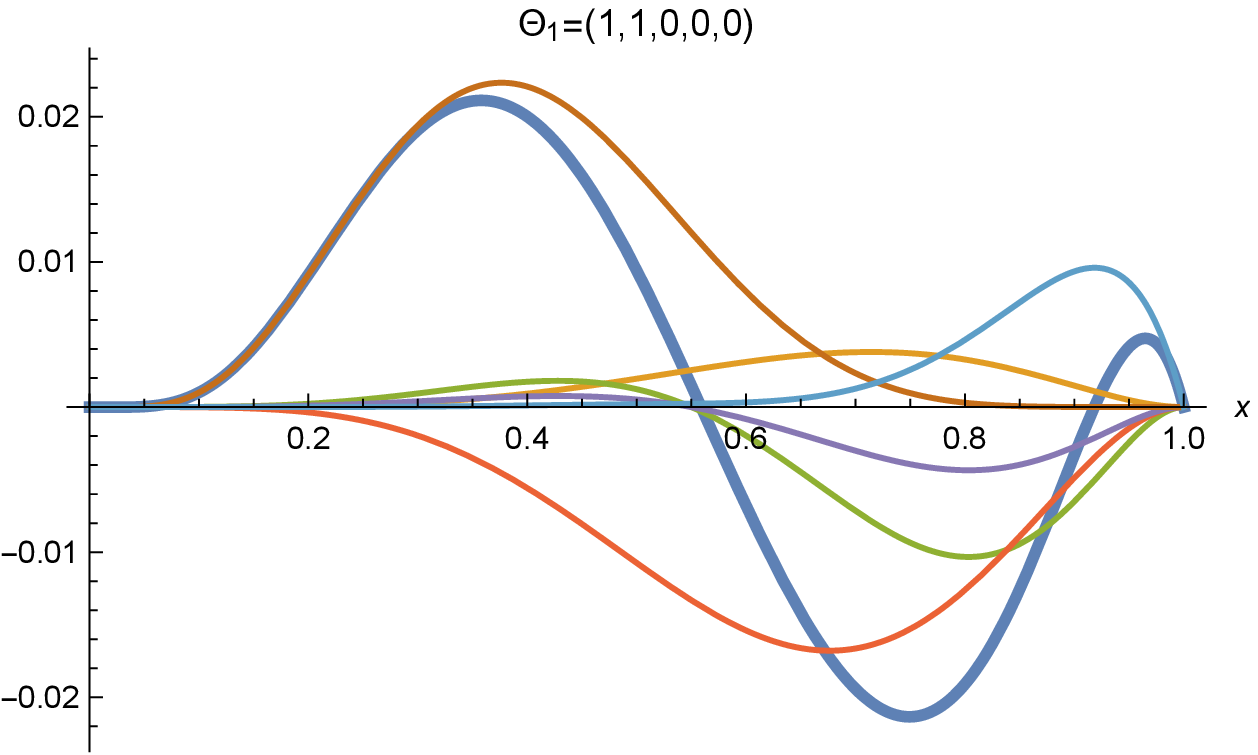}
            \caption[]%
            {{\small $g_3$}}    
        \end{subfigure}
        \hfill
        \begin{subfigure}[b]{0.45\textwidth} 
            \centering 
            \includegraphics[width=\textwidth]{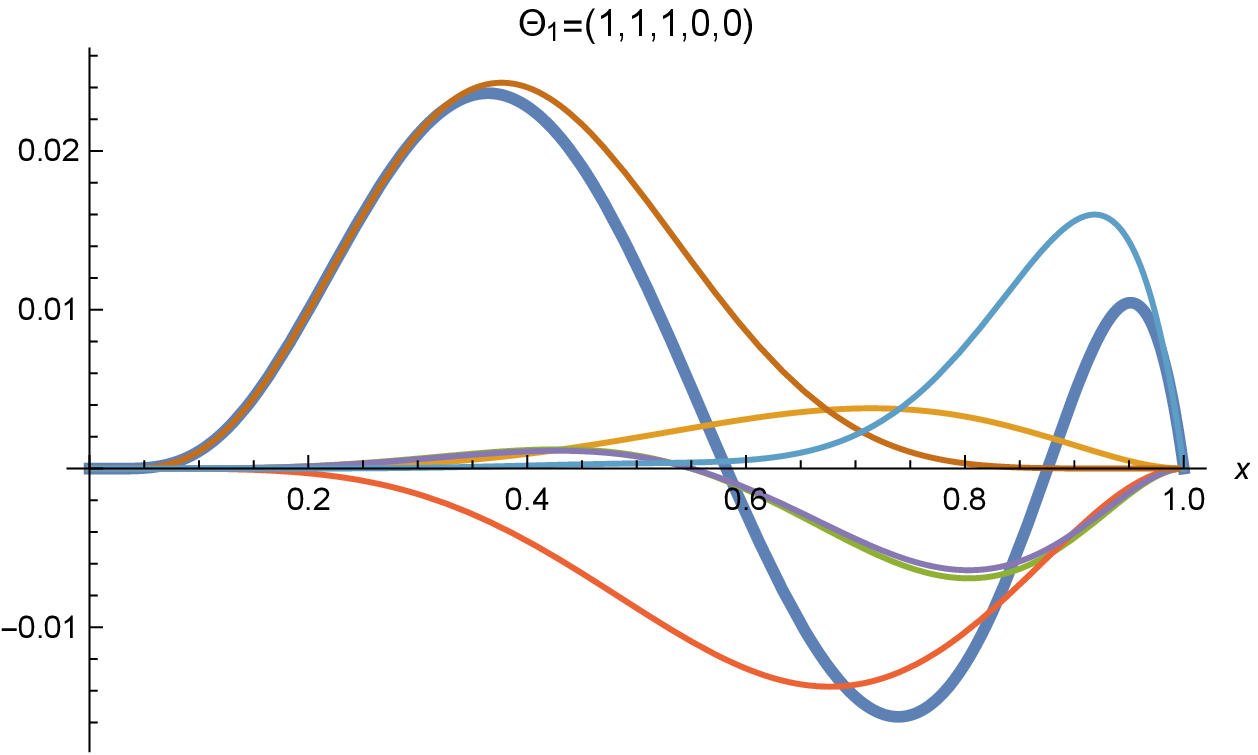}
            \caption[]%
            {{\small $g_3$}}    
        \end{subfigure}
            \hfill
        \begin{subfigure}[b]{0.45\textwidth} 
            \centering 
            \includegraphics[width=\textwidth]{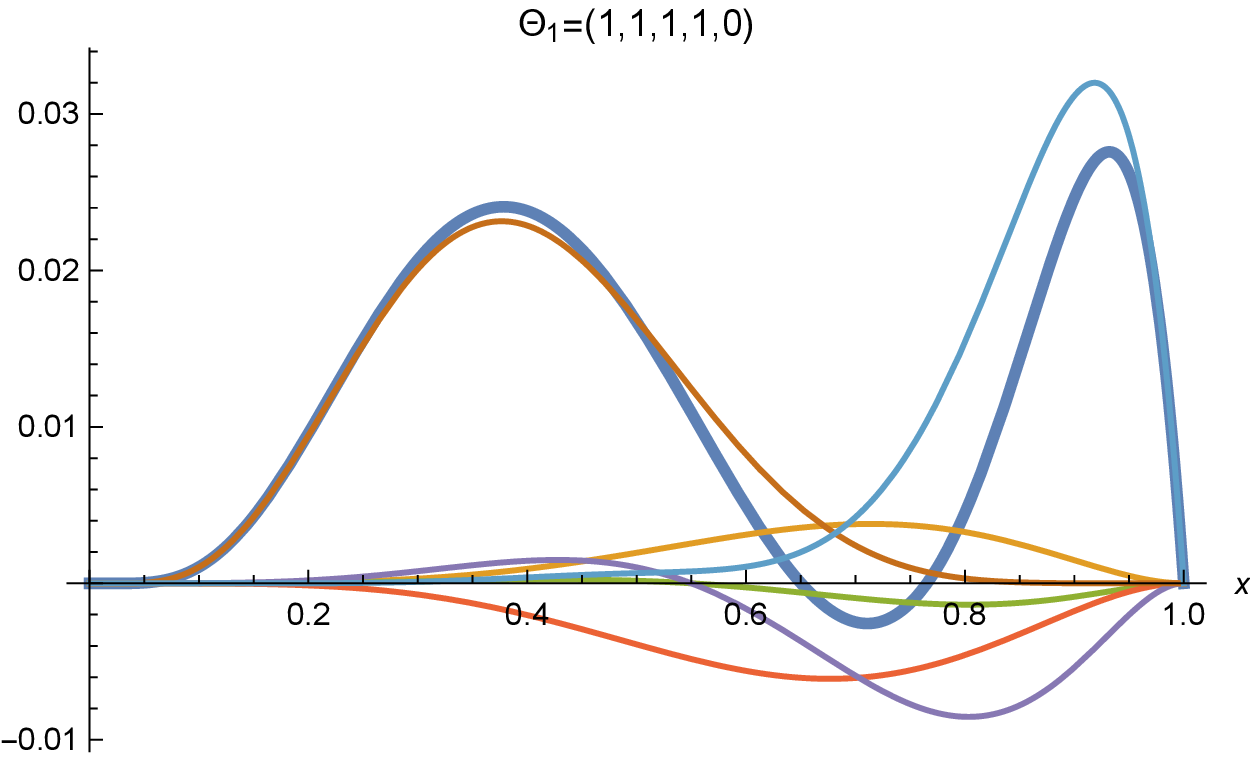}
            \caption[]%
            {{\small $g_3$}}    
        \end{subfigure}
        \hfill
        \begin{subfigure}[b]{0.45\textwidth} 
            \centering 
            \includegraphics[width=\textwidth]{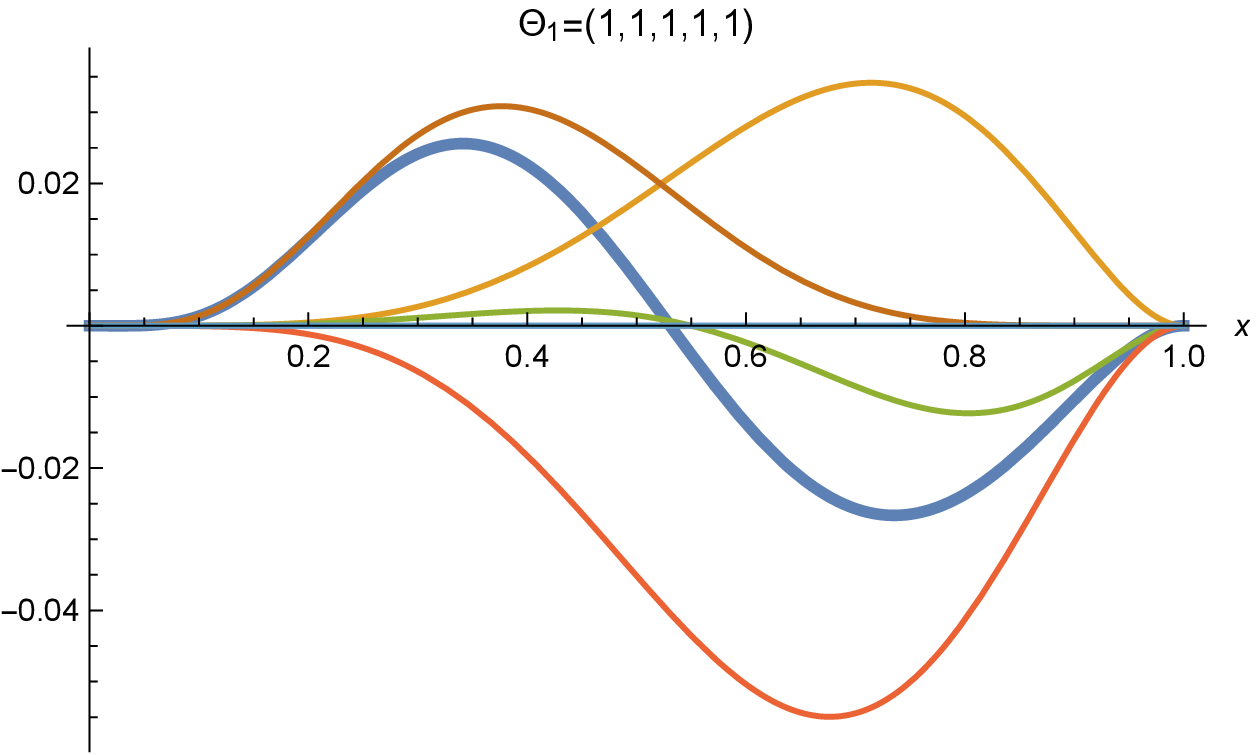}
            \caption[]%
            {{\small $g_3$}}    
        \end{subfigure}
        \caption[ ]
        {\small Full structure function $g_3$ (blue line) as a function of the Bjorken parameter $x$, obtained from the contribution of the twist-4 spin-1/2 fermionic operators ${\cal {O}}_{k=1}^{I_1, (6)}$ with $l_5=1\geq l_4\geq l_3\geq l_2\geq l_1$. We distinguish the contributions from the minimal coupling $\beta^2_m F_1^m$ (orange line); the Pauli interaction between states belong to same Kaluza-Klein tower: where the intermediate state  $\lambda_{\cal {X}} \equiv \lambda_{k=1}^-$ is the same as the incident state $\beta^2_P F_1^P$ (green line) and where the intermediate state is different  $\beta^2_{Pm} F_1^{P}$  (violet line); the contribution from crossed terms $\beta_m \beta_P F_1^c$ (red line); and the contributions from the Pauli interaction involving states which are dual to operators belonging to the $\textbf{60}$ irreducible representation of $SU(4)_R$:  $\beta^2_+ F_1^{P +}$, where $\lambda_{\cal {X}} \equiv \lambda_{k+1}^+$ (brown line), and also the $\textbf{4}$ irreducible representation of $SU(4)_R$: $\beta^2_- F_1^{P-}$, where $\lambda_{\cal {X}} \equiv \lambda_{k-1}^+$  (light blue line).}
        \label{fig-g3-detail}
    \end{figure}

\begin{figure}[h]
        \centering
        \begin{subfigure}[b]{0.45\textwidth} 
            \centering 
            \includegraphics[width=\textwidth]{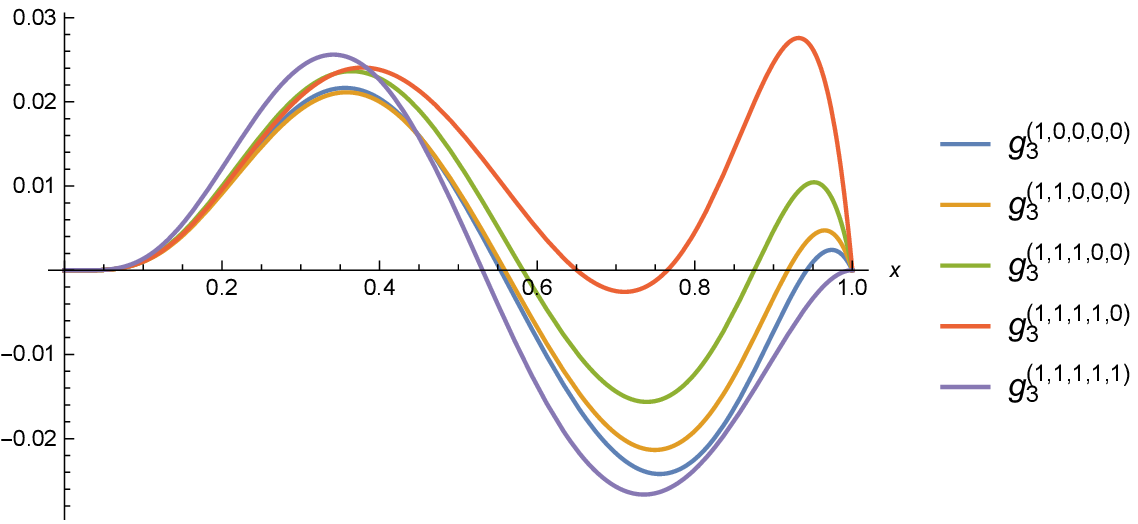}
            \caption[]%
            {{\small $g_3$}}    
        \end{subfigure}
        \hfill
        \begin{subfigure}[b]{0.45\textwidth} 
            \centering 
            \includegraphics[width=\textwidth]{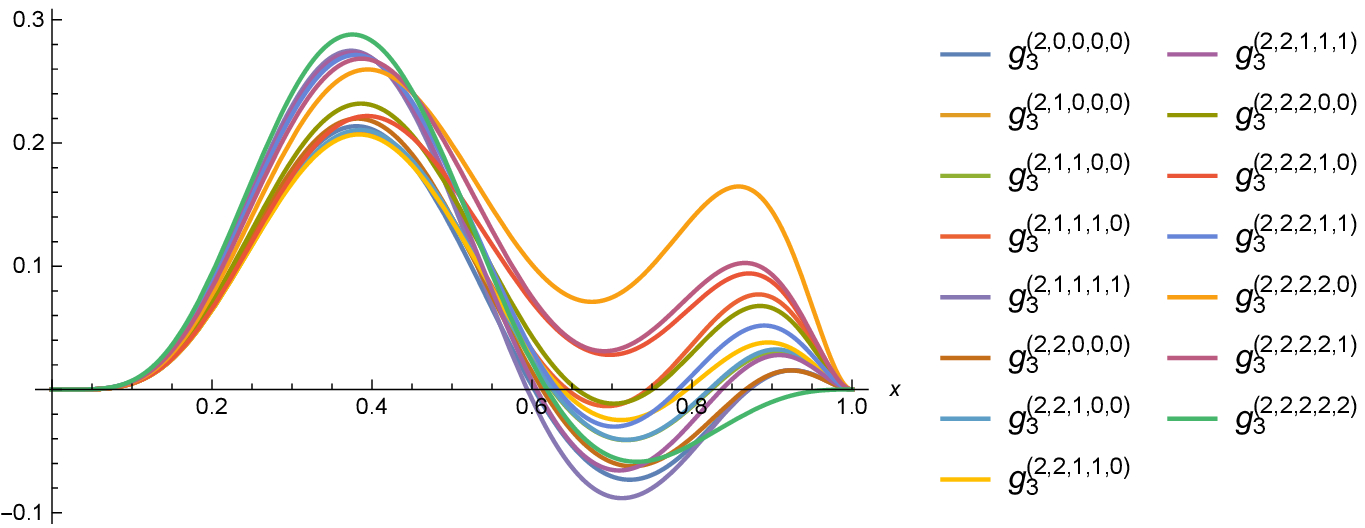}
            \caption[]%
            {{\small $g_3$}}    
        \end{subfigure}
        \caption[ ]
        {\small Full structure function $g_3$ as a function of the Bjorken parameter $x$ obtained from the twist-4 spin-1/2 fermionic operators ${\cal {O}}_{k=1}^{I_1, (6)}$ with $(1,l_4,l_3,l_2,l_1)_{a=1}$ (a) and in (b) from the twist-5 spin-1/2 fermionic operators ${\cal {O}}_{k=2}^{I_2, (6)}$ with $(2,l_4,l_3,l_2,l_1)_{a=1}$.} 
        \label{fig-g3-completo-tau-4-5}
    \end{figure}

\section{Spherical harmonics }
\label{Appendix_D}

In this appendix some of the spinor spherical harmonics used to calculate the structure functions are explicitly written. In order to build them we have employed the formalism proposed in reference \cite{Camporesi:1995fb}.     

\subsection{The case with $\tau=4$ ($l_5=1$)}

We first list the spinor spherical harmonics with ${\cal{Q}}=1/2$. Notice that $\Theta^{-}_{(1,0,0,0,0)_{a=1}}$ is given in Section 3.1 in equation (\ref{spinor10000}).
\begin{equation}
{\scriptstyle    \Theta^{-}_{(1,1,0,0,0)_{a=1}}}= 
{\scriptstyle \frac{ \sqrt{3}e^{-i{\cal{Q}} \theta_1}}{\sqrt{10}\pi^{3/2}}} 
\begin{bmatrix} {\scriptstyle e^{-i \frac{1}{2}( \theta_3 -\theta_5)} 
\cos (\frac{\theta_2}{2}) \cos (\frac{\theta_4}{2})\sin{(\theta_5)}\left(1-5\cos{(\theta_4)}\right)}  \\ 
{\scriptstyle e^{i \frac{1}{2}( \theta_3 +\theta_5)} 
\sin{ (\frac{\theta_2}{2})} \cos (\frac{\theta_4}{2})\sin{(\theta_5)}\left(-1+5\cos{(\theta_4)}\right) }\\ 
{\scriptstyle e^{-i \frac{1}{2}(\theta_3 +\theta_5)} 
\cos{ (\frac{\theta_2}{2})} \sin{(\frac{\theta_4}{2})}\sin{(\theta_5)}\left(1+5\cos{(\theta_4)}\right) }\\
{\scriptstyle e^{-i \frac{1}{2}( -\theta_3 +\theta_5)} 
\sin{(\frac{\theta_2}{2})} \sin {(\frac{\theta_4}{2})}\sin{(\theta_5)}\left(1+5\cos{(\theta_4)}\right)}  \end{bmatrix}  \, ,
\end{equation}
\begin{equation}
 {\scriptstyle   \Theta^{-}_{(1,1,1,0,0)_{a=1}}= 
\frac{ \sqrt{2}e^{-i{\cal{Q}} \theta_1}}{\pi^{3/2}} 
}\begin{bmatrix}{\scriptstyle e^{-i \frac{1}{2}( \theta_3 -\theta_5)} 
\cos (\frac{\theta_2}{2}) \cos^2(\frac{\theta_4}{2}) \sin{(\frac{\theta_4}{2})} \sin{(\theta_5)}\left(-3   i \cos{(\theta_3)}-\sin{(\theta_3)}\right)}  \\ 
{\scriptstyle e^{i \frac{1}{2}( \theta_3 \theta_5)} 
\sin (\frac{\theta_2}{2}) \cos^2(\frac{\theta_4}{2}) \sin{(\frac{\theta_4}{2})} \sin{(\theta_5)}\left(3   i \cos{(\theta_3)}+\sin{(\theta_3)}\right)}  \\ 
{\scriptstyle e^{-i \frac{1}{2}( \theta_3 +\theta_5)} 
\cos (\frac{\theta_2}{2}) \sin^2(\frac{\theta_4}{2}) \cos{(\frac{\theta_4}{2})} \sin{(\theta_5)}\left(-3   i \cos{(\theta_3)}-\sin{(\theta_3)}\right)} \\
{\scriptstyle e^{-i \frac{1}{2}( -\theta_3 +\theta_5)} 
\sin (\frac{\theta_2}{2}) \sin^2(\frac{\theta_4}{2}) \cos{(\frac{\theta_4}{2})} \sin{(\theta_5)}\left(3   i \cos{(\theta_3)}-\sin{(\theta_3)}\right)} \end{bmatrix}  \, ,
\end{equation}
\begin{equation}
  \frac{\scriptstyle \Theta^{-}_{(1,1,1,1,0)_{a=1}}}{ 
\frac{e^{-i{\cal{Q}} \theta_1}}{\pi^{3/2}}}=
 \begin{bmatrix} {\scriptstyle \sin (\theta_2) (1-3 \cos (\theta_2)) \csc \left(\frac{\theta_2}{2}\right) \sin \left(\frac{\theta_3}{2}\right) \sin (\theta_3) \left(\cot \left(\frac{\theta_3}{2}\right)-i\right) \sin ^2(\theta_4) \csc \left(\frac{\theta_4}{2}\right) \sin \left(\frac{\theta_5}{2}\right) \sin (\theta_5) \left(\cot \left(\frac{\theta_5}{2}\right)+i\right)} \\
 {\scriptstyle 2\sin \left(\frac{\theta_2}{2}\right) (3 \cos (\theta_2)+1) \sin \left(\frac{\theta_3}{2}\right) \sin
   (\theta_3) \left(\cot \left(\frac{\theta_3}{2}\right)+i\right) \sin ^2(\theta_4) \csc \left(\frac{\theta_4}{2}\right) \sin \left(\frac{\theta_5}{2}\right) \sin (\theta_5) \left(\cot \left(\frac{\theta_5}{2}\right)+i\right)} \\  
{\scriptstyle 2\sin (\theta_2) (3 \cos (\theta_2)-1) \csc \left(\frac{\theta_2}{2}\right) \sin \left(\frac{\theta_3}{2}\right) \sin (\theta_3) \left(\cot \left(\frac{\theta_3}{2}\right)-i\right) \sin \left(\frac{\theta_4}{2}\right) \sin (\theta_4) \sin \left(\frac{\theta_5}{2}\right) \sin (\theta_5) \left(\cot \left(\frac{\theta_5}{2}\right)-i\right)} \\
{\scriptstyle -4\sin \left(\frac{\theta_2}{2}\right) (3 \cos (\theta_2)+1) \sin
   \left(\frac{\theta_3}{2}\right) \sin (\theta_3) \left(\cot \left(\frac{\theta_3}{2}\right)+i\right)
   \sin \left(\frac{\theta_4}{2}\right) \sin (\theta_4) \sin \left(\frac{\theta_5}{2}\right) \sin
   (\theta_5) \left(\cot \left(\frac{\theta_5}{2}\right)-i\right) } \end{bmatrix}  \,
\end{equation}

\newpage

Next, for ${\cal{Q}}=3/2$ we have

\begin{equation}
   {\scriptstyle \Theta^{-}_{(1,1,1,1,1)_{a=1}}= 
\frac{ \sqrt{3}e^{-i{\cal{Q}} \theta_1}}{4\pi^{3/2}} } 
\begin{bmatrix}  
{\scriptstyle i e^{-\frac{1}{2} i (\theta_3-\theta_5)}\sin ^2(\theta_2) \csc \left(\frac{\theta_2}{2}\right) \sin (\theta_3) \sin^2(\theta_4) \csc \left(\frac{\theta_4}{2}\right) \sin (\theta_5) } \\
{\scriptstyle 2 i e^{\frac{1}{2} i (\theta_3+\theta_5)} \sin \left(\frac{\theta_2}{2}\right) \sin (\theta_2) \sin (\theta_3) \sin^2(\theta_4) \csc \left(\frac{\theta_4}{2}\right) \sin (\theta_5)  }\\ 
{\scriptstyle 2 i  \sin ^2(\theta_2) \csc \left(\frac{\theta_2}{2}\right) \sin
   \left(\frac{\theta_3}{2}\right) \sin (\theta_3) \left(\cot \left(\frac{\theta_3}{2}\right)-i\right)
   \sin \left(\frac{\theta_4}{2}\right) \sin (\theta_4) \sin \left(\frac{\theta_5}{2}\right) \sin
   (\theta_5) \left(\cot \left(\frac{\theta_5}{2}\right)-i\right) } \\
{\scriptstyle 4 i e^{-\frac{1}{2} i (-\theta_3+\theta_5)} \sin \left(\frac{\theta_2}{2}\right) \sin (\theta_2) \sin (\theta_3) \sin
   \left(\frac{\theta_4}{2}\right) \sin (\theta_4) \sin (\theta_5) }  \end{bmatrix}  \,  \\
\end{equation}

\subsection{The case with $\tau=5$ ($l_5=2$)}

There are ten spinor spherical harmonics with ${\cal{Q}}=1/2$, four with ${\cal{Q}}=3/2$ and only one with ${\cal{Q}}=5/2$. We only show a few examples for each charge.

For ${\cal{Q}}=1/2$ we display the spinor spherical harmonic $(2,0,0,0,0)_{a=1}$ as follows,
\begin{equation}
    {\scriptstyle \Theta^{-}_{(2,0,0,0,0)_{a=1}}= 
\frac{ \sqrt{3}e^{-i{\cal{Q}} \theta_1}}{\sqrt{5}\pi^{3/2}} } 
\begin{bmatrix} {\scriptstyle i e^{-\frac{1}{2} i (\theta_3-\theta_5)} \cos \left(\frac{\theta_2}{2}\right) \cos \left(\frac{\theta_4}{2}\right)
    (i \sin (2 \theta_5)+3 \cos (2 \theta_5)+2) } \\
 {\scriptstyle e^{-\frac{1}{2}
   i (-\theta_3-\theta_5)} \sin \left(\frac{\theta_2}{2}\right) \cos \left(\frac{\theta_4}{2}\right)(\sin (2 \theta_5)-3 i \cos (2 \theta_5)-2 i) } \\
{\scriptstyle
-i e^{-\frac{1}{2} i (\theta_3+\theta_5)}\cos \left(\frac{\theta_2}{2}\right) \sin \left(\frac{\theta_4}{2}\right) (-i \sin (2 \theta_5)+3 \cos (2 \theta_5)+2) }\\
{\scriptstyle
e^{-\frac{1}{2}
   i (-\theta_3+\theta_5)} \sin \left(\frac{\theta_2}{2}\right) \sin \left(\frac{\theta_4}{2}\right) (\sin (2 \theta_5)+3 i \cos (2 \theta_5)+2 i) }  \end{bmatrix}  \, ,
\end{equation}

\begin{equation}
   {\scriptstyle \Theta^{-}_{(2,1,
    0,0,0)_{a=1}}= 
\frac{ \sqrt{3}e^{-i{\cal{Q}} \theta_1}}{\sqrt{70}\pi^{3/2}} } 
\begin{bmatrix}{\scriptstyle i e^{-\frac{1}{2} i (\theta_3-\theta_5)} \cos \left(\frac{\theta_2}{2}\right) \cos \left(\frac{\theta_4}{2}\right) (5 \cos
   (\theta_4)-1) \sin (\theta_5) (7
   \cos (\theta_5)+i \sin (\theta_5)) }  \\ 
{\scriptstyle e^{-\frac{1}{2} i (-\theta_3-\theta_5)} \sin \left(\frac{\theta_2}{2}\right) \cos \left(\frac{\theta_4}{2}\right) (5 \cos (\theta_4)-1) \sin (\theta_5) (\sin
   (\theta_5)-7 i \cos (\theta_5))} \\ 
{\scriptstyle -e^{-\frac{1}{2} i (\theta_3+\theta_5)} \cos \left(\frac{\theta_2}{2}\right) \sin \left(\frac{\theta_4}{2}\right) (5 \cos (\theta_4)+1) \sin (\theta_5) (\sin (\theta_5)+7 i \cos (\theta_5))}
\\
{\scriptstyle e^{-\frac{1}{2} i (-\theta_3+\theta_5)}\sin \left(\frac{\theta_2}{2}\right) \sin \left(\frac{\theta_4}{2}\right) (5 \cos (\theta_4)+1) \sin (\theta_5) (\sin (\theta_5)+7 i \cos (\theta_5)) }  \end{bmatrix}  \, .
\end{equation}

For charge ${\cal{Q}}=3/2$ we write the example:

\begin{equation}
{\scriptstyle
    \frac{\Theta^{-}_{(2,2,
    1,1,1)_{1}}}{\frac{  e^{-i{\cal{Q}} \theta_1}}{\sqrt{7}\pi^{3/2}} }}= 
\begin{bmatrix} 
{\scriptstyle
4 i e^{-\frac{1}{2} i (\theta_3-\theta_5)} \sin \left(\frac{\theta_2}{2}\right) \cos ^2\left(\frac{\theta_2}{2}\right) \sin (\theta_3)
   \left(7 \sin \left(\frac{3 \theta_4}{2}\right)-9 \sin \left(\frac{\theta_4}{2}\right)\right) \cos
   ^2\left(\frac{\theta_4}{2}\right) \sin ^2(\theta_5)}
   \\ 
   {\scriptstyle
\sin \left(\frac{\theta_2}{2}\right) \sin (\theta_2) \sin
   \left(\frac{\theta_3}{2}\right) \sin (\theta_3) \left(\cot \left(\frac{\theta_3}{2}\right)+i\right)
   \sin ^2(\theta_4) (7 \cos (\theta_4)-1) \csc \left(\frac{\theta_4}{2}\right) \sin
   \left(\frac{\theta_5}{2}\right) \sin ^2(\theta_5) \left(1-i \cot \left(\frac{\theta_5}{2}\right)\right)} \\ 
{\scriptstyle
 \sin ^2(\theta_2) \csc \left(\frac{\theta_2}{2}\right) \sin
   \left(\frac{\theta_3}{2}\right) \sin (\theta_3) \left(-1-i \cot \left(\frac{\theta_3}{2}\right)\right)
   \sin \left(\frac{\theta_4}{2}\right) \sin (\theta_4) (7 \cos (\theta_4)+1) \sin
   \left(\frac{\theta_5}{2}\right) \sin ^2(\theta_5) \left(\cot \left(\frac{\theta_5}{2}\right)-i\right)}
\\
{\scriptstyle
2 \sin \left(\frac{\theta_2}{2}\right) \sin (\theta_2) \sin
   \left(\frac{\theta_3}{2}\right) \sin (\theta_3) \left(\cot \left(\frac{\theta_3}{2}\right)+i\right)
   \sin \left(\frac{\theta_4}{2}\right) \sin (\theta_4) (7 \cos (\theta_4)+1) \sin
   \left(\frac{\theta_5}{2}\right) \sin ^2(\theta_5) \left(1+i \cot \left(\frac{\theta_5}{2}\right)\right)} \end{bmatrix}  \, 
\end{equation}

Finally, for ${\cal{Q}}=5/2$ we have: 
\begin{equation}
{\scriptstyle
    \Theta^{-}_{(2,2,
    2,2,2)_{a=1}}= 
\frac{\sqrt{3}  e^{-i{\cal{Q}} \theta_1}}{\sqrt{2}\pi^{3/2}}}
\begin{bmatrix} 
{\scriptstyle
 i \frac{e^{-\frac{1}{2} i (\theta_3)}}{2} \sin ^3(\theta_2) \csc
   \left(\frac{\theta_2}{2}\right) \sin ^2(\theta_3) \sin ^3(\theta_4) \csc \left(\frac{\theta_4}{2}\right) \sin \left(\frac{\theta_5}{2}\right) \sin ^2(\theta_5) \left(\cot \left(\frac{\theta_5}{2}\right)+i\right)}
   \\ 
   {\scriptstyle
-i e^{\frac{1}{2} i (\theta_3+\theta_5)} \sqrt{\frac{3}{2}} \sin \left(\frac{\theta_2}{2}\right) \sin ^2(\theta_2) \sin ^2(\theta_3) \sin
   ^3(\theta_4) \csc \left(\frac{\theta_4}{2}\right) \sin ^2(\theta_5)} \\ 
{\scriptstyle
  \sin ^3(\theta_2) \csc \left(\frac{\theta_2}{2}\right) \sin \left(\frac{\theta_3}{2}\right) \sin ^2(\theta_3) \left(-1-i \cot \left(\frac{\theta_3}{2}\right)\right) \sin \left(\frac{\theta_4}{2}\right) \sin ^2(\theta_4) \sin \left(\frac{\theta_5}{2}\right) \sin ^2(\theta_5) \left(\cot \left(\frac{\theta_5}{2}\right)-i\right)}
\\
{\scriptstyle
2i e^{\frac{1}{2} i (\theta_3-\theta_5)} \sqrt{6} \sin \left(\frac{\theta_2}{2}\right) \sin ^2(\theta_2) \sin ^2(\theta_3) \sin
   \left(\frac{\theta_4}{2}\right) \sin ^2(\theta_4) \sin ^2(\theta_5)}
   \end{bmatrix}  \, 
\end{equation}

\end{document}